\documentclass[fleqn,usenatbib]{mnras}

\usepackage{newtxtext,newtxmath}
\usepackage{tabularx}
\usepackage[normalem]{ulem}
\usepackage{hyperref}
\usepackage{subfig}
\usepackage{threeparttable}

\usepackage[T1]{fontenc}

\DeclareRobustCommand{\VAN}[3]{#2}
\let\VANthebibliography\thebibliography
\def\thebibliography{\DeclareRobustCommand{\VAN}[3]{##3}\VANthebibliography}

\usepackage{subcaption}
\usepackage{adjustbox}
\usepackage{graphicx}	
\usepackage{amsmath}	
\usepackage{threeparttable}

\usepackage{xcolor}

\usepackage{cleveref} 

\title[SN explosion models and iPTF16geu]{iPTF16geu through the lens of thermonuclear explosion models} 

\author[A. Sainz de Murieta et al.]{
Ana Sainz de Murieta$^{1}$\thanks{E-mail: ana.sainz-de-murieta1@port.ac.uk},
Mark R. Magee$^{2}$, Tian Li$^{1}$, Thomas E. Collett$^{1}$, Joel Johansson$^{3}$
\\
$^{1}$Institute of Cosmology and Gravitation, University of Portsmouth, Burnaby Road, Portsmouth, PO1 3FX, UK\\
$^{2}$Department of Physics, University of Warwick, Gibbet Hill Road, Coventry CV4 7AL, UK\\
$^{3}$Department of Physics, The Oskar Klein Center, Stockholm University, AlbaNova, 10691 Stockholm, Sweden
}

\date{Accepted 2026 January 04. Received 2025 December 17; in original form 2025 October 01}

\pubyear{2025}

\begin{document}
\label{firstpage}
\pagerange{\pageref{firstpage}--\pageref{lastpage}}
\maketitle

\begin{abstract} 
The magnification resulting from strong gravitational lensing is a powerful tool to add new constraints to the cosmic evolution of supernova progenitors by enabling the study of distant supernovae that would otherwise not be observable. iPTF16geu is the most well-observed gravitationally lensed supernova (glSN) to date. At a redshift of $z = 0.409$ and magnified by a factor of $\sim$68, extensive photometric and spectroscopic observations have been obtained. The explosion mechanism producing this rare event and differences compared to lower redshift supernovae however have not been explored in detail. Here we compare observations of iPTF16geu to existing radiative transfer simulations of type Ia supernova explosion models selected from the literature. We find that overall the DDC6, PDDEL1, and N10 models produce the closest match to the light curves and many absorption features, providing some evidence in favour of the delayed detonation scenario. All models struggle however to replicate the observed colours and in particular the rest-frame UV. We also investigate the magnification and reddening values required to improve agreement with the selected models. Upcoming surveys will significantly increase the samples of SNe discovered at high redshifts due to strong gravitational lensing. These glSNe will enable tighter constraints on the explosion physics of type Ia supernovae and how this has evolved throughout the Universe. 
\end{abstract}

%


\section{Introduction}
Type Ia supernovae (SNe~Ia) played a key role in the discovery of the accelerating expansion rate of the Universe \citep{Ries1998, Perlmutter} due to their standardisable nature. Their absolute magnitudes can be corrected using empirical relationships, allowing them to be used as precise distance indicators. 

Standardisation with SNe~Ia typically follows the Tripp equation \citep{tripp98} and accounts for correlations between the supernova peak brightness, light curve shape, and colour. Applying the same corrections to SNe~Ia across a range of redshifts however implicitly assumes that SNe~Ia properties do not evolve significantly with redshift. Nevertheless, some evolution may be expected. The differences in stellar birth environments with redshift (e.g., \citealt{LeFloch05}) will affect the progenitor and explosion mechanism \citep{Rigault2013, Childress14} and potentially introduce additional systematics \citep{Linder2006}.

Multiple explosion and progenitor scenarios have been proposed to explain the observational properties of SNe~Ia (see \citealt{ruiter--25} for a recent review). Discriminating between them is non-trivial, but the ultraviolet (UV) region of the spectrum contains valuable information about the explosion physics and progenitor systems \citep{Lentz, Hoeflich}. These wavelengths are strongly affected by line-blanketing due to iron-group elements and therefore are ideal for probing the metal content of the progenitor and SN \citep{Bufano2009}. Indeed, population studies of SNe~Ia in the UV have found significantly more diversity than observed at optical wavelengths \citep{Foley2016} and evidence of a systematic UV flux excess at intermediate redshifts ($z\sim0.5$; \citealt{Ellis2008, Maguire2012, Foley2012}). This may be indicative of differences in the metallicity of the progenitors compared to those at lower redshifts \citep{Foley2008}.

Despite the utility of these high-redshift SNe, current samples are limited by both the quality and quantity of UV spectra available -- typically only single epochs for individual SNe~Ia (e.g., \citealt{Foley_2008, brown2018}). 
Therefore even individual high signal-to-noise ratio (S/N) events at high redshift are valuable for studying deviations from the trends observed in large, low-redshift samples. The difficulties in observing these high-redshift SNe~Ia in the UV can be overcome via strong gravitational lensing. Strong gravitational lensing boosts the apparent brightness of SNe and can significantly improve the S/N. Additionally, most discoverable gravitationally lensed supernovae (glSNe) are expected to be found around $z\sim0.8-1$ \citep{Goldstein19, SainzdeMurieta23}, where the UV rest-frame shifts into the optical range, and can therefore be probed through observer-frame optical photometry. While the current sample of glSNe is small, the Vera Rubin Observatory's Legacy Survey of Space and Time (LSST; \citealt{lsst}) is expected to detect hundreds of glSNe over its 10-year survey \citep{Goldstein19,Arendse24, SainzdeMurieta23, SainzdeMurieta24}.

In this work, we focus on the glSN iPTF16geu \citep{goobar--17}. \cite{goobar--17} present the discovery and first study of iPTF16geu, showing it matches the template of a normal SN~Ia at $z = 0.409$ magnified by a factor of $\gtrsim$50. Further spectroscopic analysis and comparisons with low redshift SNe~Ia by \cite{Cano2018} and \cite{Johansson2020} revealed iPTF16geu to be a high velocity and high velocity gradient SN~Ia \citep{benetti--05, branch--06, wang--09}. \cite{Gall24} investigate the strong sodium absorption observed in iPTF16geu and find that this is most likely due to interstellar dust rather than circumstellar material surrounding the supernova. Observations of iPTF16geu have also been used for studies on lens modelling \citep{Mortsell:2019auy}, dust \citep{Dhawan2019}, and microlensing \citep{Diego2022, Arendse25}. To date however, iPTF16geu has yet to be studied in the context of explosion scenarios and their redshift evolution.

Previous studies of other glSNe have also focused on attempts to constrain their progenitor scenarios. \cite{Petrushevska17} analysed early-time data from PS1-10afx, a glSN at $z = 1.39$, and found no evidence of redshift-dependent spectral differences with respect to local SNe. More recently, \cite{Dhawan2024} analysed JWST data from SN~Encore, a glSN~Ia at $z = 1.95$, and also found no significant redshift evolution. These studies were however limited by the range of phases covered by the spectra of both objects. Conversely, iPTF16geu has a large amount of photometric and spectroscopic data covering up to two rest-frame months after peak, providing a valuable opportunity to compare against different explosion and progenitor scenarios.

This work provides the first comparison of spectroscopic and photometric glSN data to explosion models, showcasing the potential of glSNe as an astrophysical probe. This paper is structured as follows. We present a summary of the iPTF16geu data used throughout this work in \Cref{sec:observations}. In \Cref{sec:models} we describe the literature explosion models used in our analysis. \Cref{sec:analysis} presents comparisons between iPTF16geu and our selected models, assuming the magnification and extinction values calculated by \cite{Dhawan2019}, while in \Cref{sec:new_dust_estimates} we calculate the best-fitting magnification and extinction parameters for each model. We discuss our results in \Cref{sec:discussion} and summarise our conclusions in \Cref{sec:conclusions}.

\section{Observations of iPTF16geu}
\label{sec:observations}

Throughout this work, we use the unresolved $griz$-band light curves of iPTF16geu presented by \cite{goobar--17} and \cite{Dhawan2019}, which range from 12\,d pre-maximum to 49\,d post-maximum. Following from \cite{Dhawan2019}, we assume a total lensing magnification of $\mu_{\mathrm{TOT}} = 67.8 ^{+2.6} _ {-2.9}$ and reddening values of  $R_{V, \mathrm{host}} = R_{V, \mathrm{lens}} = 2$. Based on resolved HST and Keck images, \cite{Dhawan2019} calculate an extinction correction for each lensed image. Here we combine these in a single, magnification-weighted extinction value of $E(B-V)_{\mathrm{lens}}$ = 0.34 for the unresolved light curves. In \Cref{sec:new_dust_estimates} we explore the impact of different magnifications and reddening parameters for each model. Assuming a flat $\Lambda$CDM cosmology with $H_0=67.66$ km s$^{-1}$, $\Omega_m$ = 0.31 \citep{planck18--20}, we calculate a distance modulus of $\mu =$ 41.8 mag.

Table \ref{tab:16GEU} gives details of the spectra used throughout our analysis. This includes early-time spectra presented by \cite{goobar--17}, GTC+OSIRIS and VLT+XSHOOTER spectra presented by \cite{Cano2018}, and additional spectra from Keck, the Discovery Channel Telescope (DCT), and the Palomar 200-inch Hale Telescope (P200) presented by \cite{Johansson2020}. Spectroscopy of the lens and host galaxies after the supernova had faded was taken with the P200 telescope and is used as reference for galaxy fitting. 

\begin{table}
	\centering
	\caption{Log of spectroscopic observations of iPTF16geu used in this analysis. The last spectrum was taken $\approx3$ years after the SN.}
	\label{tab:16GEU}
	\begin{tabular}{lcccc} 
		\hline
		\textbf{UT date} & \textbf{MJD} & \textbf{Phase} & \textbf{Telescope} & \textbf{Source}\\
		\hline
		2016-10-02.23 & 57663.23 & +7.4 & P60+SEDM & G17\\
		2016-10-04.22 & 57665.22 & +8.8  & P200+DBSP & G17\\
		2016-10-06.13 & 57665.13 & +10.2 & P200+DBSP & G17\\
        2016-10-09.90 & 57670.90 & +12.8 & NOT+ALFOSC & G17\\
        2016-10-15.87 & 57676.87 & +17.1 & GTC + OSIRIS & C18\\
        2016-10-22.34 &  57681.34 &+20.3 & DCT+DeVeny & J20\\
        2016-10-25.33 & 57686.33 &  +23.8 & Keck+DEIMOS & J20\\
        2016-10-26.19 & 57687.10  & +24.3 & P200+DBSP & J20\\
        2016-10-30.01& 57691.01 & +27.1 & VLT+XSHOOTER & C18\\
        2016-11-02.24 & 57694.24 & +29.4 & Keck+LRIS & J20\\
        2016-11-18.86 & 57710.86 & +41.2 & GTC+OSIRIS & C18\\
        2016-11-28.21 & 57720.21 & +47.9 & Keck+LRIS & J20\\
        2016-11-30.82 &  57723.82 & +49.7 & GTC+OSIRIS & C18\\
        2016-12-15.81 & 57737.81 & +60.3 & GTC+OSIRIS & C18\\
         \hline
         \textbf{2019-05-24.46} & \textbf{58626.96} & \textbf{--} & \textbf{P200+DBSP} & \textbf{J20}\\
	\hline
	\end{tabular}
\begin{tablenotes}\footnotesize
\item G17: \cite{goobar--17}, C18: \cite{Cano2018}, J20: \cite{Johansson2020}
\end{tablenotes}
\end{table}

\section{Models}
\label{sec:models}

Here we give a brief overview of the explosion models used in this paper. We refer the reader to the referenced sources for more detailed descriptions of each model. Our model sample includes pure detonations, double detonations, pure deflagrations, and delayed detonations. In this work we focus on specific models within each category that were found to produce the closest agreement to the peak magnitude and shape of the light curves of iPTF16geu. Light curve parameters for the models used are given in~\Cref{tab:models}.

\subsection{Pure detonations}
\cite{Blondin17} present models invoking the detonation of a sub-Chandrasekhar mass white dwarf (WD). In these models, the centre of a C/O WD, consisting of equal amounts of carbon and oxygen, is artificially ignited and the one-dimensional explosion is calculated using the method detailed by \cite{Khokhlov1991}. The resulting ejecta are then used as input for \textsc{cmfgen} \citep{Hillier98, Hillier12}, a one-dimensional, time-dependent radiative transfer code that assumes non-local thermal equilibrium (NLTE) and spherical symmetry to calculate synthetic light curves and spectra. These models have characteristically high $^{56}$Ni-mass-to-total-mass ratios, which results in high temperatures and ionization levels. Compared to, for example, delayed detonation models synthesizing a similar amount of $^{56}$Ni, pure detonations result in bluer colours and lower amounts of line blanketing near maximum light. These differences are greatly reduced a few weeks after maximum. \cite{Blondin17} show that these models find reasonable agreement with the fainter end of the Phillips relation \citep{1993ApJ...413L.105P}, similar to \cite{Sim2010}. Here we focus on the SCH7p0 model\footnote{Available at \url{https://zenodo.org/records/8379254}}, which results from the explosion of a 1.15 $M_{\odot}$ mass WD and predicts 0.84 $M_{\odot}$ of $^{56}$Ni.

\subsection{Double detonations}

\cite{doubledetonations2021} show a range of double detonation models, in which a helium shell is ignited on the surface of an accreting C/O WD. The three-dimensional simulations for these models are computed using the moving mesh code \textsc{arepo} \citep{arepo}. In these models, the He detonations can be ignited through compressional heating of the accreted material. This detonation propagates through the He shell and sends a shock wave into the C/O core \citep{Fink2010}. A second detonation in the C/O material is ignited following the convergence of the shock. Burning products from the He shell detonation include significant amounts of iron-group elements and, as such, have a strong impact on bluer wavelengths of the spectrum. 

\cite{collins22} use the \textsc{artis} three-dimensional radiative transfer code \citep{ARTIS1, ARTIS2} to calculate synthetic observables for these models. \cite{collins22} show that the light curves broadly reproduce the faint end of the width–luminosity relation, but due to line blanketing caused by the He-shell ash the models tend to be redder than normal SNe~Ia. This is similar to previous studies of double detonations \citep{kromer, woosleykasen, gronow20,kenshen}. These models also show a clear Ti absorption feature between 4\,000 -- 4\,500~\AA, which is characteristic of some sub-luminous SNe~Ia (e.g., \citealt{filippenko--92}).

For this work, we use the M11\_05 model\footnote{This spectrum as well as other double detonation models are available on \url{https://zenodo.org/records/7997388}}, which consists of a C/O WD of 1.1 $M_{\odot}$ and a 0.054 $M_{\odot}$ He shell. This model produces 0.83 $M_{\odot}$ of $^{56}$Ni from the core detonation and 1.2 $\times 10^{-2} M_{\odot}$ of $^{56}$Ni from the shell detonation. \cite{collins22} find strong viewing angle dependencies for this model, which we comment on in Section \ref{sec:discussion}. Here we focus only on the angle-averaged spectra. 

\subsection{Pure deflagrations}
Pure deflagration models of Chandrasekhar mass ($M_{\textrm{Ch}}$) C/O WDs are presented by \cite{fink--14}. These models are parameterized by the number of sparks used to ignite the deflagration phase. The hydrodynamic simulations are carried out in three dimensions with the finite volume hydrodynamics code \textsc{leafs} \citep{LEAFS} and synthetic observables are calculated using time-dependent three-dimensional Monte Carlo radiative transfer simulations with the\textsc{artis} code. Pure deflagrations produce small $^{56}$Ni masses and fail to reproduce observations of normal SNe~Ia, but have shown promise for the fainter SNe~Iax class. Although generally not suitable for normal SNe~Ia, we include pure deflagrations within our analysis due to the high levels of mixing predicted. This results in a significantly different distribution of iron-group elements than expected for normal SNe~Ia and therefore serves as a useful reference point for the impact on the UV. Here we include the N150def pure deflagration model from \cite{fink--14}, which synthesizes 0.378 $M_{\odot}$ of $^{56}$Ni.

\subsection{Delayed detonations}
In this scenario, a $M_{\mathrm{Ch}}$ C/O WD experiences an initial deflagration phase before transitioning to a supersonic detonation \citep{Khokhlov1991}. The exact details of the transition from deflagration to detonation are not well-constrained; therefore, we include three subtypes of models within the delayed detonation framework.

\cite{Seitenzahl2013} present a suite of delayed detonation (DDT) models invoking various strengths for the deflagration phase. The strength of the deflagration is determined by the number of sparks used to ignite the model. Models with a larger number of sparks exhibit a stronger deflagration phase and therefore produce less $^{56}$Ni due to the commensurate decrease in density during the detonation phase. Three-dimensional hydrodynamic simulations were calculated using \textsc{leafs}. Radiative transfer simulations calculated with \textsc{artis} are presented by \cite{Sim2013}. Here we use the N10 model, which has 10 ignition sparks, resulting in 0.93 $M_{\odot}$ of $^{56}$Ni. \cite{Sim2013} point out that the optical light curves for their DDT models appear to decline slightly more quickly post-maximum than do the bulk of normal SNe Ia, but they generally produce reasonable spectroscopic agreement with observations of normal SNe~Ia.

We also include comparisons with the Gravitationally Confined Detonation (GCD) scenario presented by \cite{Lach22GCD}. In these models, after the weak deflagration, the deflagration burns to the surface of the white dwarf and wraps around the gravitationally bound core. As the deflagration ash collides on the opposite side of the breakout, a detonation is ignited that leads to an explosion disrupting the entire white dwarf. The hydrodynamic simulations are also carried out using \textsc{leafs} and the synthetic observables calculated using \textsc{artis}. These models display different ejecta structures compared to other delayed detonations due to the highly asymmetric and non-central nature of the beginning of the detonation phase, and due to it beginning much later in the explosion. The deflagration ashes extend out to high ejecta velocities and surround the detonation products. As opposed to the pure deflagrations, they are able to reproduce luminosities closer to normal SNe~Ia, but spectroscopically show reasonable agreement with over-luminous and slow-declining 91T-like supernovae \citep{Lach22GCD}. Here we include the r51\_d4 model, which starts with a central density of 4$\times$10$^9$~g~cm$^{-3}$ and produces 1.057 $M_{\odot}$ of $^{56}$Ni.

In both of these cases, the ashes of the deflagration (which contain the majority of the neutron-rich, stable IGE) can buoyantly rise before the detonation starts. The detonation then burns the rest of the white dwarf almost instantaneously. This results in the deflagration ashes being found outside the detonation ashes. This is a unavoidable consequence of three-dimensional delayed detonation models \citep{Pakmor24}. In one-dimensional models however, this buoyancy cannot be treated and therefore deflagration ashes become trapped in the centre with the detonation ashes on the outside. These two configurations lead to very different ejecta structures, and hence observables, for models calculated in one dimension compared to multiple. To compare between these cases, we also include the one-dimensional DDC6 model from \cite{blondin--13}, which is calculated using the same method previously described for pure detonations by \cite{Blondin17}. This model has a slightly lower central density than the N10 model from \cite{Seitenzahl2013} (2.6$\times$10$^9$~g~cm$^{-3}$ c.f. 2.9$\times$10$^9$~g~cm$^{-3}$) and produces a slightly smaller $^{56}$Ni mass of 0.73 $M_{\odot}$. Synthetic observables are calculated using \textsc{cmfgen}. As with \cite{Sim2013}, \cite{blondin--13} find reasonable agreement between their delayed detonation models and observations of normal SNe~Ia.

\cite{Dessart14} present a slightly modified delayed detonation scenario -- pulsational-delayed detonations (PDDEL). Following modest expansion of the WD during the deflagration phase, nuclear burning stops and material begins to infall. A detonation is then triggered. Compared to the previously described delayed detonation scenario, this scenario results in a greater amount of unburned carbon after the explosion. Synthetic observables for these models are also simulated with \textsc{cmfgen}. \cite{Dessart14} show that these models produce observables that are generally consistent with those of other delayed detonation models, but they exhibit bluer colours and higher luminosities at earlier epochs. Here we consider the PDDEL1 model, which produces 0.76 $M_{\odot}$ of $^{56}$Ni\footnote{Both DDC and PDDEL model spectra are available at \url{https://www-n.oca.eu/supernova/snia/snia_ddc_pddel.html}.}. Despite differences in the ignition of the detonation, the DDC and PDDEL models both result in similar observables around maximum light.

\section{Analysis}
\label{sec:analysis}

\subsection{Photometric comparison}
\label{sec:photometry}

\begin{figure}
    \centering
    \includegraphics[width=0.5\textwidth]{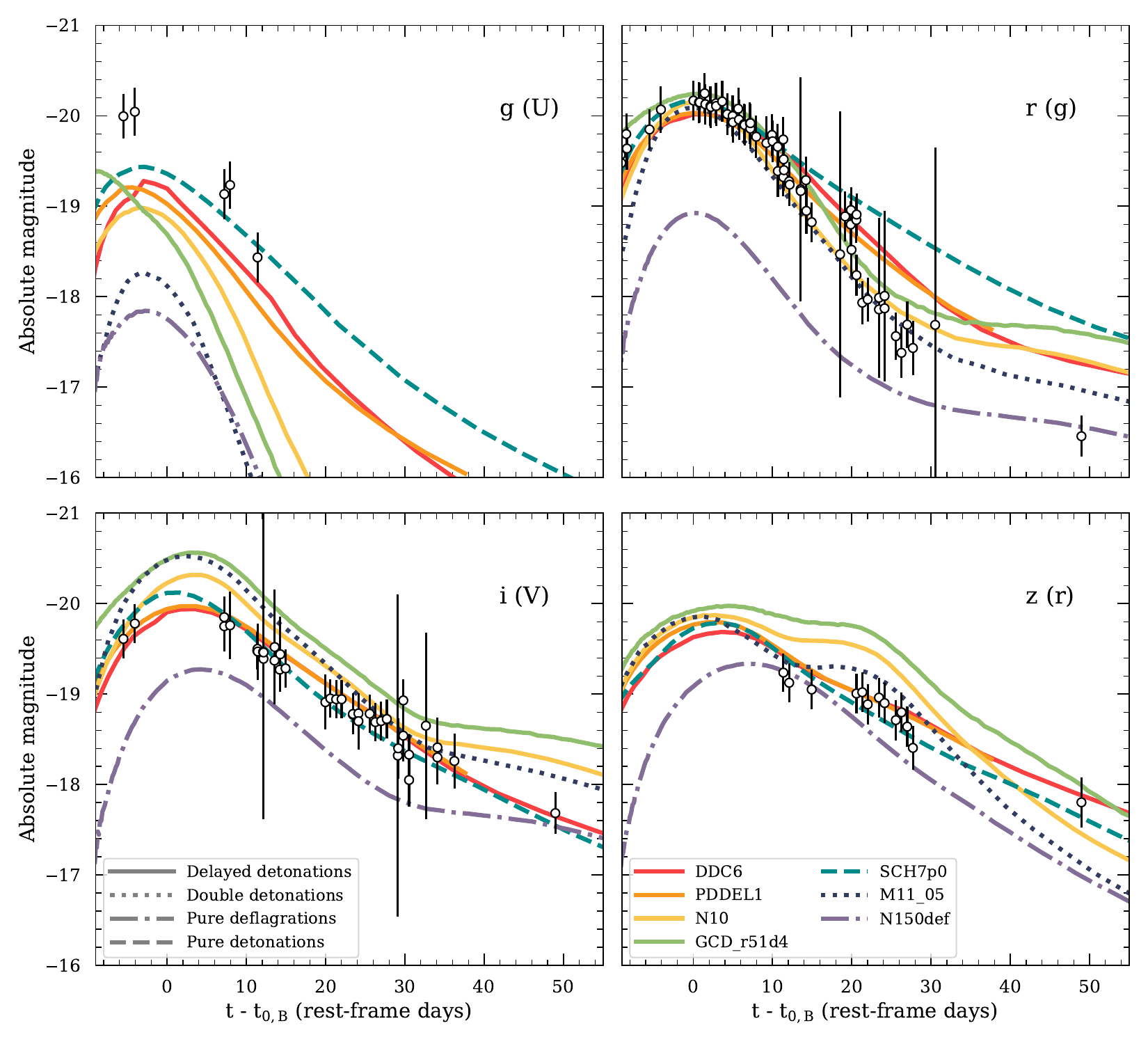}
    \caption{Light curves of iPTF16geu compared against different explosion models. Each panel shows the observer-frame band and its corresponding approximate rest-frame band in brackets. Delayed detonation models are represented by the solid lines, pure detonations are represented by dashed lines, double detonations by a dotted line, and pure deflagrations by a dash-dotted line. Times are shown with respect to peak $B-$band.}
    \label{fig:norm_lc}
\end{figure}

In \Cref{fig:norm_lc}, we show the light curves of iPTF16geu compared against our models transformed to the observer frame of the SN ($z = 0.409$). Observed light curves of iPTF16geu have been corrected for the lensing magnification and extinction values presented by \cite{Dhawan2019}. These values were calculated assuming a normal SNe Ia template \citep{hsiao2007}, but the uncertainty in magnification and the effect of $K$-corrections make the light curves appear slightly brighter compared to most normal SNe~Ia. In \Cref{sec:new_dust_estimates} we calculate the specific magnification and extinction parameters for each model that best reproduce the observations. The peak magnitudes and decline rates in each band for our models and iPTF16geu are given in \Cref{tab:models}. For those bands without observations around peak, we fit the light curve using the SALT3 template assuming a fixed time of maximum in the $B$-band from \cite{goobar--17}. The error bars account for systematic errors arising from observations, lensing magnification estimates, and extinction uncertainties, as well as statistical errors from the template fitting.

The $r$-band is the best sampled band for iPTF16geu and corresponds to an effective wavelength of $\sim4\,400$~\AA\, in the rest-frame, which is close to the rest-frame $g$-band. We find the peak magnitude in this band to be $M_r$ = $-20.2\pm0.2$. As expected, the N150def model is significantly fainter than iPTF16geu, with $M_r$ = $-18.9$. All other models produce similar $r$-band peak absolute magnitudes to iPTF16geu within the uncertainty. The $r$-band light curve of iPTF16geu has a decline rate of $\Delta m_{15,r} = 0.60\pm 0.11$, which is slower than predicted by the N10 model ($\Delta m_{15,r} = 0.84$). The DDC6, PDDEL1, GCD\_r51d4 and SCH7p0 models all predict broadly similar decline rates in agreement with iPTF16geu ($\Delta m_{15,r}$ = 0.49 -- 0.65). While the N150def deflagration model shows a faster  decline ($\Delta m_{15,r} = 0.81$), we note that this model also predicts the presence of a luminous, bound remnant after the explosion. The inclusion of this remnant in radiative transfer simulations slows the overall light curve evolution and may produce decline rates more similar to iPTF16geu \citep{Kromer2013, Callan}. The M11\_05 model also exhibits a faster decline than iPTF16geu in this band, with $\Delta m_{15,r}$ = 0.84.

\par

The observer-frame $i$-band corresponds to approximately the rest-frame $V$-band. Here, we estimate a peak absolute magnitude of $M_i = -19.9 \pm 0.2$ and decline rate of $\Delta m_{15,i} = 0.44\pm 0.10$ for iPTF16geu. This is in agreement with the DDC6 and PDDEL1 models. The SCH7p0 pure detonation model also has a comparable $i$-band magnitude ($M_i = -20.1$), but declines faster ($\Delta m_{15,i} = 0.61$). The N10, GCD\_r5d14, and M11\_05 models are slightly brighter than iPTF16geu ($M_i = -20.3$, $-20.6$ and $-20.5$, respectively) and decline faster ($\Delta m_{15,i} = 0.69$, 0.61 and 0.63, respectively). In this case, the N150def model shows a comparable decline rate ($\Delta m_{15,i} = 0.49$), but remains 0.6~mag fainter at peak.

\par

The $z$-band (corresponding approximately to a rest frame of $r$-band) light curve begins only approximately 10 days post-maximum. The peak magnitude is therefore somewhat uncertain and dependent on our \textsc{sncosmo} fit, but we estimate $M_z \approx -19.8\pm 0.4$ and $\Delta m_{15,z}=0.60\pm 0.07$. With the exception of N150def, all other models are consistent with the estimated peak magnitude of iPTF16geu. All delayed detonation models however, have much slower decline rates ranging from $\Delta m_{15,z} = 0.19$ (GCD\_r51d4) to $\Delta m_{15,z} = 0.41$ (DDC6). The M11\_05 double detonation model shows a slightly slower decline of $\Delta m_{15,z} = 0.51$, while the pure detonation model is within 1$\sigma$ of our measurements for iPTF16geu ($\Delta m_{15,z} = 0.54$). The N150def model remains fainter than iPTF16geu by 0.5~mag and also has a slower decline ($\Delta m_{15,z} = 0.42$). We note that iPTF16geu may show some sign of a weak secondary maximum beginning around a phase of $\gtrsim$20\,d, although the relatively large uncertainties make it difficult to fully constrain its presence. The N10 and GCD\_r51d4 delayed detonation models and the double detonation model (M11\_05) also show secondary maxima, with the latter being somewhat weaker and marginally closer to observations of iPTF16geu (but see Sect.~\ref{sec:new_dust_estimates}).

\par

The observer-frame $g$-band corresponds to a rest-frame effective wavelength of $\sim$3\,300~\AA\ (which is slightly bluer than the rest-frame $U$-band) and therefore allows us to investigate the near-UV (NUV) properties of iPTF16geu. \Cref{fig:norm_lc} shows that iPTF16geu is brighter than predictions from all models, with a peak magnitude of $M_g = -20.0\pm 0.2$. The SCH7p0 pure detonation model has the brightest peak $g$-band magnitude, but is still $\textgreater$0.5~mag fainter than iPTF16geu, with an absolute magnitude $M_g = -19.4$. The overall shape of the $g$-band light curve however shows improved agreement. For iPTF16geu, we find a decline rate of $\Delta m_{15,g} = 0.71\pm 0.16$, which is consistent with the SCH7p0, DDC6, PDDEL1, and GCD\_r51d4 models. The N10 model declines slightly faster than iPTF16geu ($\Delta m_{15,g} = 0.98$), while the N150def and M11\_05 models are both significantly faster ($\Delta m_{15,g} \textgreater 1.05$).

\begin{figure}
   \centering
    \includegraphics[width = 0.5\textwidth]{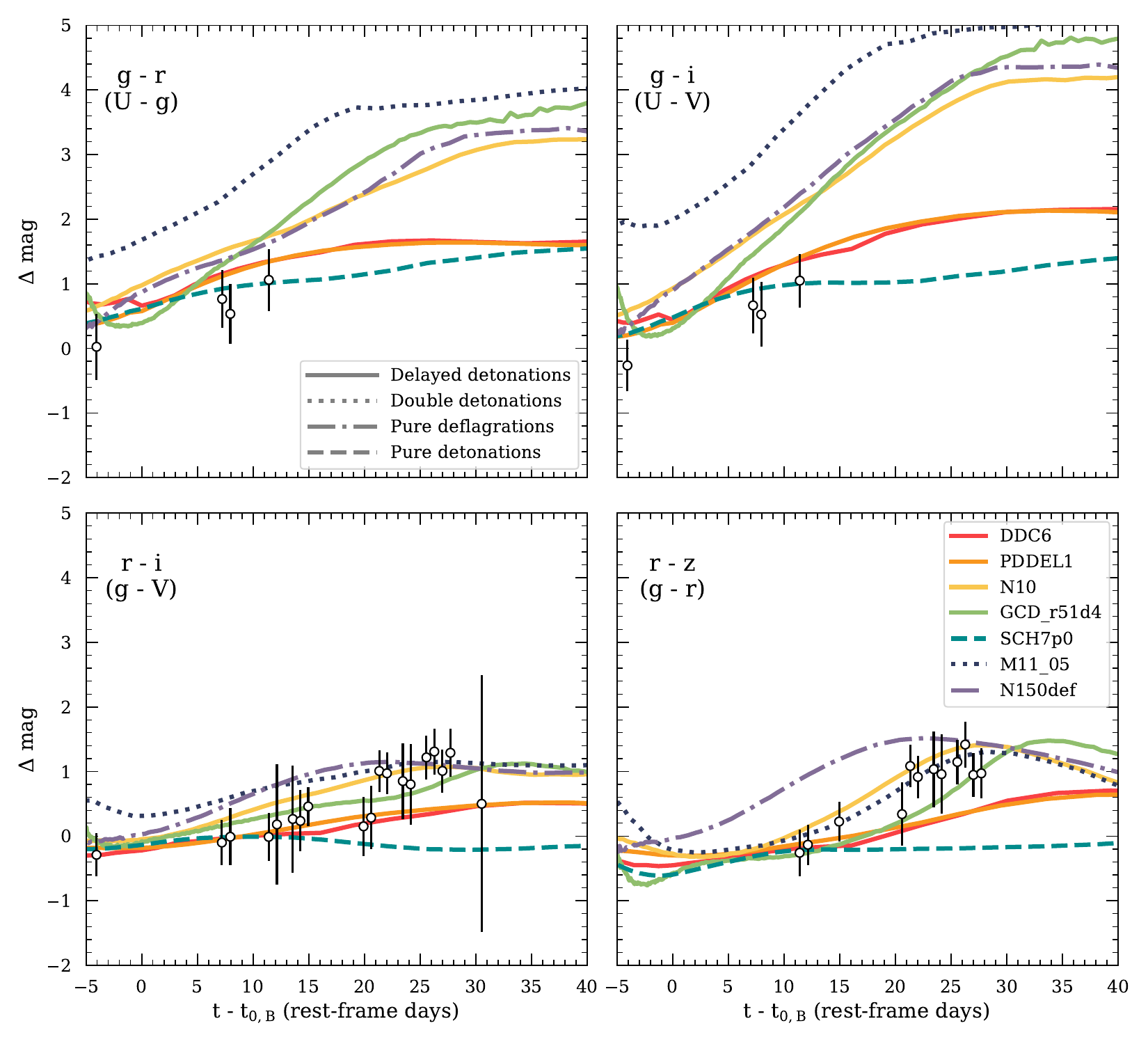}
    \caption{Colour evolution of iPTF16geu and our comparison models. Each panel shows the observer-frame colour and its corresponding rest-frame band in brackets. Times are in reference to the $B$-band peak. Observed colours of iPTF16geu have been corrected for extinction.}
    \label{fig:colourcurves}
\end{figure}

Based on the light curves shown in \Cref{fig:norm_lc}, it is clear that no single model matches the colours or evolution observed in iPTF16geu. This is further confirmed in \Cref{fig:colourcurves}, which shows the colour evolution of iPTF16geu and our comparison models. The $g-r$ colour (approximately rest-frame $U-g$) of iPTF16geu evolves from $g-r = -0.15 \pm 0.42$~mag at 5 days before $r$-band peak to $g-r = 1.06 \pm 0.48$~mag at 11 days after $r$-band peak, resulting in a colour slope of $\sim$0.08$\pm$0.04~mag~d$^{-1}$.
Given the relatively large uncertainties, the SCH7p0 pure detonation model shows an approximately comparable colour evolution, although it is somewhat redder before peak. The N150def model is redder and also has a faster colour gradient than iPTF16geu. The N10 and M11\_05 models show similar slopes to those observed in iPTF16geu, but are systematically redder by $\sim$0.6 -- 1.8~mag. The DDC6 and PDDEL1 models are also generally consistent with the data and show a similar colour evolution, with the exception of times earlier than $B$-band peak where the DDC6 model is redder by $\sim$0.3~mag. We find a similar trend in the $g-i$ (rest frame $U-V$) colours. The N150def model becomes redder at a faster rate than iPTF16geu, while the M11\_05 and N10 models show a consistent colour evolution and are systematically redder. Again, the DDC6, PDDEL1, and SCH7p0 models are consistent with the observed colours of iPTF16geu in these bands.

Conversely, the $r-i$ colours (approximately rest-frame $g-V$) of iPTF16geu have much stronger similarities to the models, which also display less variation. iPTF16geu shows an increase in colour of $\sim$0.014$\pm$0.03~mag~d$^{-1}$, from $-0.23\pm0.37$~mag at $-5$\,d to $-0.01\pm0.42$~mag at $+11$\,d. 
The pure detonation (SCH7p0) and delayed detonation models show a similar evolution at early ($t \lesssim +10$\,d) phases. After approximately $+$10 -- 20\,d however, these models predict a relatively flat colour evolution while iPTF16geu continues to get redder. Again, we note that the large uncertainties mean that most models (with the exception of the SCH7p0 pure detonation model) are consistent within $\sim$1 -- 2$\sigma$ of the observed colours at these phases. For the $r-z$ colour (approximately rest-frame $g-r$), the PDDEL1 and DDC6 models again show qualitatively good agreement, but are somewhat too blue at later times. Likewise, the SCH7p0 pure detonation model is too blue beginning at $\sim+$20\,d, which roughly corresponds with the beginning of the possible secondary maximum in iPTF16geu.
We note that the colours of iPTF16geu are also consistent with a flat colour slope with time, which would indicate no colour evolution. This is however not expected for SNe~Ia, and we conclude that this represents the limitations of our data rather than an intrinsic property of iPTF16geu.

In summary, we find that qualitatively the DDC6, PDDEL1 and N10 models show the strongest similarities with iPTF16geu. These models generally provide a reasonable agreement with the observer-frame $r$- and $i$-bands, but exhibit slower declines in the $z$-band and fainter $g$-band magnitudes. The SCH7p0 pure detonation model also reproduces some properties, such as the $ri$-band magnitudes and the shape of the $g$-band light curve, but fails to reproduce the observed colours in the redder bands. Further implications of these results are discussed in \Cref{sec:discussion}.

\subsection{Spectroscopic comparison}
\label{sec:spectroscopy}

\begin{figure*}
    \centering
    \includegraphics[width=\textwidth]{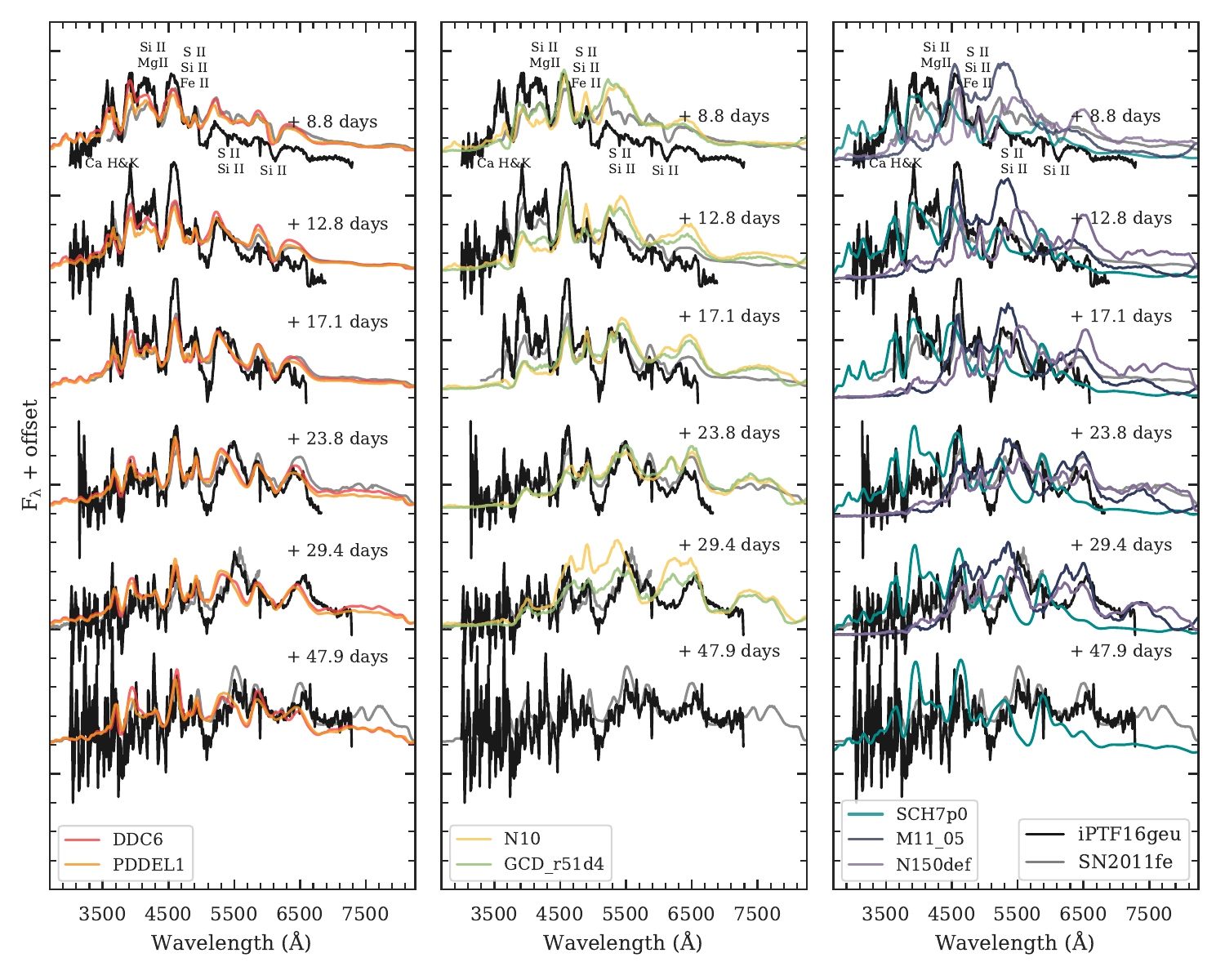}
    \caption{Spectra of iPTF16geu and comparison models at different phases relative to $B$-band maximum. Spectra are normalised by the median flux in the wavelength region shown and offset for clarity. Spectra of iPTF16geu have been corrected for extinction. The grey comparison spectra belong to SN2011fe.}
    \label{fig:spectra}
\end{figure*}

As demonstrated by \cite{Cano2018}, spectral features in glSNe may be strongly contaminated by light from both the lens and host galaxies. The UV in particular can be highly affected by such contamination. Therefore, robust comparisons rely on an accurate removal of the lens and host galaxy light. While the host of iPTF16geu was observed approximately 2.5 years after the SN, this P200 spectrum does not extend as far into the UV as other spectra considered here. Following \cite{Johansson2020}, we model the SEDs of both the host and lens galaxies, and subtract these from the observed spectra of iPTF16geu. In this work, we use updated galaxy models from \textsc{bagpipes} \citep{BAGPIPES}. In Appendix~\ref{appendix:galaxymodel}, we further discuss our host galaxy models and removal. In Fig.~\ref{fig:spectra} we show the spectral evolution of iPTF16geu, corrected for host and lens contamination, along with model spectra at similar phases\footnote{We do not include the classification spectrum at $+7.4$\,d due to its low resolution.}. 

Using the spectra at $+8.8$\,d and $+12.8$\,d, we find a \ion{Si}{ii}~$\lambda$6\,355\, velocity gradient of $\dot{v}_{\mathrm{Si 6355}} = -160 \pm 40$~km~s$^{-1}$~d$^{-1}$, which is consistent with the value reported by \cite{Cano2018} ($\dot{v}_{\mathrm{Si 6355}} = -110.3 \pm 10.0$~km~s$^{-1}$~d$^{-1}$) and higher than the value from \cite{Johansson2020} ($\dot{v}_{\mathrm{Si 6355}} = -82 \pm 13$~km~s$^{-1}$~d$^{-1}$). This places iPTF16geu in the high velocity gradient (HVG) group of SNe~Ia according to the \cite{2005ApJ...623.1011B} classification scheme, which is also in agreement with \cite{Cano2018}. Extrapolating our linear fit back to maximum light, we determine that $v_{\mathrm{Si 6355}} = 12\,452 \pm 391$~km~s$^{-1}$. This falls within in the high velocity class of \cite{Wang2009} and \cite{Folatelli2013}, agreeing with both \cite{Cano2018} and \cite{Johansson2020}. We perform similar fits on our models and find that only the SCH7p0 model displays a velocity gradient consistent with iPTF16geu, with $\dot{v}_{\mathrm{Si 6355}} = -114\pm9$~km~s$^{-1}$~d$^{-1}$. The \ion{Si}{ii}~$\lambda$6\,355 velocity at peak for this model however (16\,100~km~s$^{-1}$) is significantly higher. No models simultaneously match the \ion{Si}{ii} velocity at peak and its evolution. The PDDEL1 model exhibits lower \ion{Si}{ii} velocities ($v_{\mathrm{Si 6355}} = 11\,600$~km~s$^{-1}$), but the evolution is fairly flat. The DDC6, N10 and GCD\_r51d4 models show higher velocities ($\sim 13\,300$ -- $13\,400$~km~s$^{-1}$) and evolve more slowly ($\dot{v}_{\mathrm{Si 6355}}$ =  $-45$ for the first one and $\sim$ $-88$~km~s$^{-1}$~d$^{-1}$, respectively). The N150def model shows a much lower \ion{Si}{ii} velocity (7\,600 km s$^{-1}$) and evolves more rapidly.

The optical spectra of iPTF16geu exhibit a broad and complex feature between $\sim$5\,300 -- 6\,000~\AA\, corresponding to a combination of \ion{S}{ii} and \ion{Si}{ii} absorption features. A similar feature appears in the spectra of the delayed detonation models, but is much broader after $+$17.1\,d. The two broad features between 4\,500 -- 5\,300~\AA\, also contain a mix of \ion{S}{ii}, \ion{Si}{ii}, with additional \ion{Fe}{ii}, and are similarly reproduced by the delayed detonation models up to $\lesssim$20\,d after peak. Following this phase, these two features follow the same evolution in the DDC6 and PDDEL1 models as iPTF16geu, but become weaker in the N10 model. %
All other models also exhibit a similar "W"-shaped double feature, but follow a different evolution from iPTF16geu, becoming much less prominent after $+20$\,d. Greater variation is expected at these phases, because as early as 20 days post-maximum the spectra of SNe~Ia begin to transition from being dominated by P Cygni lines to being dominated by emission lines \citep{Jefferey92}.

Around 4\,500~\AA, iPTF16geu shows a strong absorption feature due to \ion{Si}{ii}~$\lambda$4\,450 and \ion{Mg}{ii}~$\lambda$4\,300. A similar feature is observed in the DDC6 and PDDEL1 delayed detonation models. The N10 and GCD\_r51d4 models also show a somewhat similar feature at $+8.8$\,d that disappears in later spectra due to the decreasing flux at wavelengths $\lesssim$4\,500~\AA.

iPTF16geu shows a strong \ion{Ca}{ii}~H\&K feature with a velocity of $14\,140 \pm 140$~km~s$^{-1}$ at $+8.8$\,d. The SCH7p0 pure detonation model shows somewhat lower velocities of $\sim$13\,800 ~km~s$^{-1}$ while the PDDEL1 model has a significantly lower velocity ($\sim$13\,100~km~s$^{-1}$). The DDC6 model shows a higher velocity of 15\,010~km~s$^{-1}$. While the model predicts a persistent feature at phases $\textgreater$30\,d, the low signal-to-noise ratio in iPTF16geu spectra at these epochs prevents further direct comparisons and robust constraints.

In Fig.~\ref{fig:spectra} we also include SN~2011fe \citep{Nugent2011} as a further point of comparison. Both SNe show similar optical features, particularly at early phases, and similarities with the DDC6 and PDDEL1 models. We find that at phases earlier than 20 days post-maximum, SN~2011fe shows better agreement with the models. The peak of the \ion{Si}{ii} and \ion{Mg}{ii} profiles around 4\,500~\AA\, is less pronounced than in iPTF16geu, matching the DDC6 and PDDEL1 model predictions. Beyond 20 days however, these models do not completely reproduce the properties of SN~2011fe. The spectra beyond $\approx$ 29 days of both SN~2011fe and iPTF16geu show almost identical properties. At these phases, the models predict very different features beyond $\approx$ 5\,500~\AA\, to the ones observed in the SNe.

For SN~2011fe, we measure a \ion{Si}{ii} velocity of $v_{\mathrm{Si 6355}} = 10\,080\pm30$~km~s$^{-1}$ around peak, consistent with a normal SN~Ia and with the value measured by \cite{pereira13}. At $+8.8$\,d, SN~2011fe shows somewhat lower \ion{Si}{ii}~$\lambda$6\,355 and  \ion{Ca}{ii}~H\&K velocities of $9\,750\pm30$~km~s$^{-1}$ and 12\,960$\pm$110~km~s$^{-1}$, respectively, compared to iPTF16geu and most of our comparison models. These velocities are in better agreement with the PDDEL1 model.

Overall, we find the PDDEL1 and DDC6 delayed detonation models provide the best agreement with the optical spectra of iPTF16geu, but the SCH7p0 pure detonation model is a closer match to the observed velocities. In general, the M11\_05 double detonation and N150def pure deflagration models show very different spectral features to those observed in iPTF16geu. No model is able to reproduce all observed features, particularly at later phases. Similar deviations are also seen in the spectra of SN~2011fe, indicating this is not specific to iPTF16geu.

\subsection{Near-ultraviolet comparison}

\begin{figure*}
   \centering
    \includegraphics[width = \textwidth]{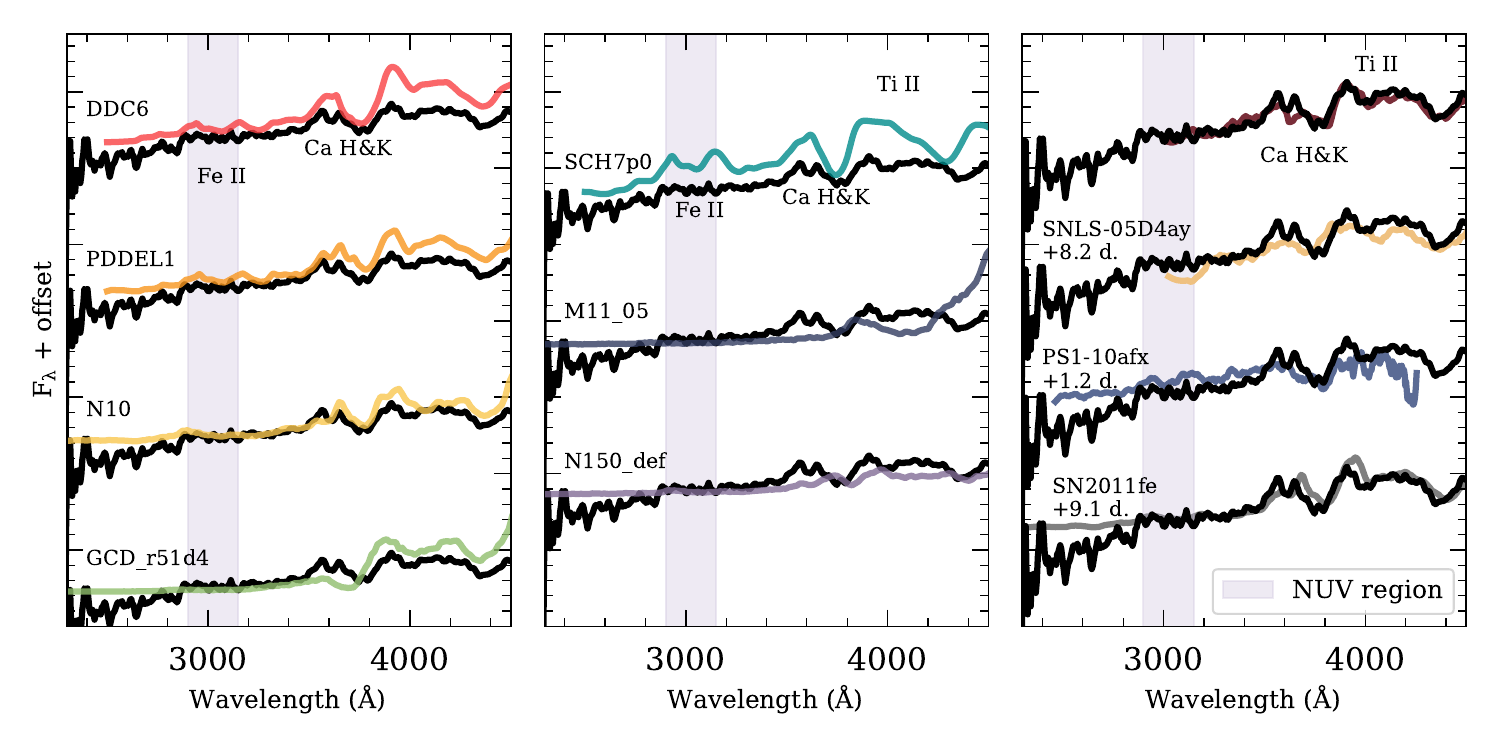}
    \caption{Zoom-in around the NUV region for iPTF16geu (black) and the explosion models and comparison SNe at phase $\approx$ +8.8 days after maximum $B$-band brigthness. The shaded region corresponds to the NUV region.}
    \label{fig:UVzoom}
\end{figure*}
 Figure~\ref{fig:UVzoom} shows a zoom-in of the UV region and the differences between each of the models. The M11\_05 model spectrum shows the most significant differences compared to iPTF16geu. The relatively large He shell results in significant line blanketing of the UV spectra and as such few discernible features are visible. This spectrum does not exhibit the characteristic \ion{Ca}{ii}~H\&K feature present in most SNe~Ia, and instead displays a strong \ion{Ti}{ii} feature centred around $\sim4\,100$~\AA. As discussed by \cite{collins22}, this feature is observed in some peculiar 91bg-like SNe~Ia and is a result of the lower temperatures and levels of ionization of the ejecta \citep{Mazzali97}. In the case of this model, this feature is also due to the large amount of Ti produced by the He shell detonation \cite{collins22} and the relatively high mass of the WD within the sub-$M_{\mathrm{Ch}}$ class \citep{Fink2010} compared to other models of the same type. The N10 delayed detonation model may exhibit hints of this feature, but they are much less pronounced. 
 
 For most models, the few features present are weaker than those observed in iPTF16geu. The PDDEL1 model shows some evidence in favour of an additional high-velocity component to the \ion{Ca}{ii}~H\&K absorption feature, but the quality of the iPTF16geu spectra makes it difficult to determine whether it shows a similar feature. The SCH7p0 pure detonation model shows some absorption features around 3\,000~\AA, possibly corresponding to \ion{Fe}{ii} 3\,250~\AA, that are not observed in iPTF16geu. This model also has a slightly broader \ion{Ca}{ii}~H\&K feature compared to iPTF16geu and does not show the secondary smaller feature centred around 3\,600~\AA. Moreover, the prominent \ion{Fe}{ii} feature around 3\,200~\AA\, is not present in the spectrum of iPTF16geu. The N150def model clearly shows lower velocities than observed in iPTF16geu, but even shifting the spectrum in velocity space, the spectral features overall do not reproduce what is observed.
  
For comparison, we include the host-subtracted spectra of unlensed SNe~Ia at similar redshifts to iPTF16geu ($z \sim 0.40$ -- 0.42) from the Supernova Legacy Survey (SNLS; \citealt{SNLS2010}): SNLS-03D1ar and SNLS-05D4ay \citep{Balland2009}. We also include the spectrum of PS1-10afx \citep{PS1_discovery}, a glSN~Ia at a redshift of $z = 1.39$ \citep{PS1_lensed}. Following the same method as for iPTF16geu, we subtract the host galaxy spectrum from \cite{PS1_discovery} and match our host-subtracted spectrum to the observed colours. We find a range of \ion{Ca}{ii}~H\&K velocities, with SNLS-03D1ar showing a slower velocity, PS1-10afx a similar velocity, and SNLS-05D4ay a higher velocity. Moreover, the \ion{Fe}{ii} 3\,250~\AA\, feature present in the SCH7p0 pure detonation model spectra is present in the PS1-10afx spectrum. Overall, this model shows reasonably good agreement with PS1-10afx. The spectrum of PS1-10afx also shows a higher UV flux, but this is likely due to the earlier phase and more rapid evolution of the UV during those phases (see, for example, \citealt{Hook2005}).

While these spectra highlight some of the diversity in UV observations of SNe~Ia, characterising specific features remains challenging as a result of the low signal-to-noise ratio in this region. Due to the difficulties in obtaining high-quality UV observations, \cite{Foley_2008} propose measuring the ratio of the flux density at two different wavelengths in order to study the UV. This method enables the UV spectral slope to be characterised while largely avoiding the effect of specific spectral features. \cite{Foley_2008} find that the UV ratio around peak is highly correlated with the luminosity of the supernova, and propose it as a potential luminosity indicator that could reduce scatter in the Hubble diagram, although confirming this would require larger population studies. To characterize the UV slope of iPTF16geu, we follow the method outlined by \cite{Foley2016} and generate a smoothed spectrum of each SN using an inverse-variance Gaussian filter \citep{blondin2006} and scale the spectra to have roughly the same flux at 4\,000~\AA. Here we consider the flux in the NUV region to be defined by $2\,900 \leq \lambda \leq 3\,150$~\AA\, following \cite{Foley2016} (see Fig.~\ref{fig:UVzoom}). The NUV flux ratio is then given as the ratio of 10 times the flux in this region divided by the median flux at $\approx$ 4000\AA\ (3\,950 -- 4\,050 \AA\, region, following \citealt{Foley2016}), hereafter $R_{\mathrm{NUV}}$. 

\par

\begin{figure}
    \centering
    \includegraphics[width=0.5\textwidth]{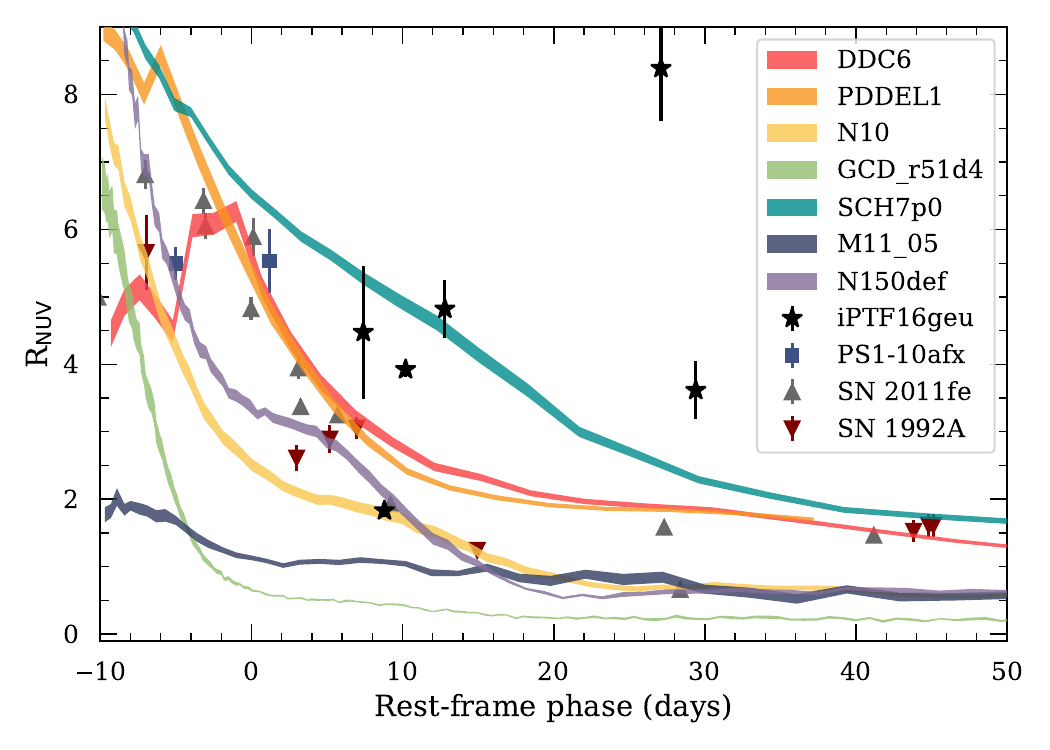}
    \caption{Evolution of the NUV flux ratio for iPTF16geu, our comparison models, and some comparison SNe (SN~2011fe, PS1-10afx and SN~1992A).}
    \label{fig:UVflux}
\end{figure}

Figure~\ref{fig:UVflux} shows $R_{\mathrm{NUV}}$ for iPTF16geu, all comparison models, other SNe~Ia (SN~1992A, SN~2011fe, and PS1-10afx). The errorbars indicate standard errors calculated via bootstrap resampling (see \citealt{Foley2008} and references within), and account for the errors in both the NUV region and the comparison region (3\,950 -- 4\,050 \AA) using error propagation. In general, the models follow an approximately exponential decay in UV flux that flattens shortly after maximum light; however, the onset and rate of this decay vary significantly between scenarios. Among the delayed detonation models, the PDDEL1 and N10 models share a similar decay rate, though PDDEL1 maintains a consistently higher UV ratio. After maximum, the DDC6 model behaves like the PDDEL1 model, but is characterized by a slower initial rise to peak UV ratio that lasts until peak $B-$band magnitude. These differences are caused by the absence of a pulsating phase in the DDC6 model which leads to different spectral compositions before maximum brightness. In contrast, the GCD\_r51d4 model exhibits a pre-peak UV ratio slightly lower than the N10 model that quickly drops to the lowest observed value ($R_{\mathrm{NUV}} \approx 0.25$), marking the fastest decay in the sample. This is because of the different composition of the GCD models, with the IGE-rich deflagration ashes in the outer layer. In contrast, the pure detonation model SCH7p0 retains a high ratio ($R_{\mathrm{NUV}}\gtrsim 3$) even at late phases, and displays the slowest decay. Finally, the N150def model shows an initially steep decline before flattening below $R_{\mathrm{NUV}}=1$, while M11\_05 model shows a mostly flat $R_{\mathrm{NUV}}$ throughout its evolution, slightly dropping from $\sim$2 -- 0.5.

The evolution of the $R_{\textrm{NUV}}$ for SN~2011fe is well-reproduced by the PDDEL1 model (as shown in Figure~\ref{fig:UVflux}). Notably, both SN~2011fe and SN~1992A exhibit $R_{\mathrm{NUV}}$ values that are consistently lower than those observed for iPTF16geu. Both of these SNe exhibit a similar UV evolution at late phases, with some scatter that can be attributed to observational effects, indicating our errorbars are probably underestimated. We also include the lensed SN PS1-10afx, which only has two spectra that extend far enough into the UV to determine the flux ratio. Both spectra occurred during the early phase of its evolution and show closer similarities to the DDC6 model rather than PDDEL1. As demonstrated by Fig.~\ref{fig:UVflux}, the models show significant diversity during the phases before and around $B$-band maximum. Obtaining observations of glSNe~Ia as early as possible will therefore be crucial in constraining their UV properties and similarities with explosion model predictions.

Our observations of iPTF16geu begin at a phase of only $+7.4$\,d and therefore we cannot constrain its evolution during the earliest epochs where the models show the most significant differences. At the observed phases, the data does not follow any of the models and also shows some scatter. In general, the observed $R_{\mathrm{NUV}}$ is higher than all of the models considered here. We note however that the signal-to-noise ratio of the iPTF16geu spectra at these phases and wavelengths is relatively low, making it difficult to fully constrain $R_{\mathrm{NUV}}$. Therefore it is not possible to determine whether the observed offset is intrinsic to iPTF16geu or due to the low quality of the spectra, or indeed other effects such as poor galaxy subtraction (see Section \ref{sec:discussion}). 

 \par

\section{Alternative magnification and extinction corrections}
\label{sec:new_dust_estimates}
Throughout the analysis presented in \Cref{sec:analysis}, we used the magnification and extinction values calculated by \cite{Dhawan2019}, which were based on assuming a \texttt{hsiao} template \citep{hsiao2007} with an additional stretch factor. This assumption introduces a bias into our analysis in which the comparison models are not necessarily directly compared to iPTF16geu and must also match the stretched template. To assess the impact of this assumption and provide a more direct comparison with the models, we fit each model to the observed data, without correcting for extinction, using \texttt{sncosmo} and determine the best-fitting magnification and extinction parameters. As our models are physically motivated by explosion and radiative transfer simulations, we do not include any additional factors, such as an additional stretch term. To reduce degeneracies between parameters we calculate a single value for the extinction, $E(B-V)_{\mathrm{lens + host}}$, applied at the redshift of the SN and present fits for two sets of $R_V$ values, $R_V$ = 2.0 and 3.1. Table~\ref{tab:magnification_dust} shows the best-fit results for each model and $R_V$ while the fitted light curves are shown in Fig.~\ref{fig:lcs_2} for the case of $R_V = 2$. Figures~\ref{fig:lcs_3} and \ref{fig:colours_3} in Appendix~\ref{appendix:rv} show the $R_V = 3.1$ fits.

The different models considered here require a wide range of magnifications to reproduce the observed light curves of iPTF16geu, varying between $20.1$ (GCD\_r51d4, $R_V = 2$) and $316.5$ (SCH7p0, $R_V = 3.1$). As expected for the $R_V  = 3.1$ case, the models require a higher magnification to compensate the greater flux loss due to extinction. In some cases (the SCH7p0 model and the DDC6 model in the $R_V=3.1$ case), the required magnification is most likely unphysical or at least highly unlikely, as it would require a lower lens mass slope than what is constrained by the velocity dispersion measurements \citep{Mortsell:2019auy}. Similarly, low magnifications (<20) would require a higher mass slope than the upper bound. We note however that the observed flux ratios for this system show the effect of microlensing by stars in the lens galaxy \citep{Dhawan2019}. Accounting for this would also affect the lens model predictions (see \citealt{Arendse25}). The predicted $E(B-V)_{\mathrm{lens + host}}$ values also vary widely depending on the model, ranging from 0.07 to 0.87 in the most extreme case.

Although some models prefer extreme magnifications, we find that our fits generally favour lower magnifications and lower (total) extinction values compared to the \texttt{hsiao} template fit by \cite{Dhawan2019}. This likely arises from the selection of relatively luminous SNe~Ia models. An exception is the SCH7p0 model, which favours both high magnification and $E(B-V)$. This model shows relatively good agreement with the light curves obtained using our fitted values and those from \cite{Dhawan2019}, highlighting the strong degeneracies between the magnification and extinction. 
\begin{table}
    \centering
	\caption{Magnification and dust extinction best-fit parameters for two different choices of $R_V$}
	\label{tab:magnification_dust}
    \resizebox{\columnwidth}{!}{
	\begin{tabular}{lcc} 		
        \hline
        Model & Magnification ($\mu$) & $E(B-V)_{\mathrm{host + lens}}$\\
        \hline
        $R_{V, \mathrm{host + lens}} = 2$&&\\
        \texttt{hsiao} \citep{Dhawan2019} & 67.8 $^{+2.5}_{-2.9}$ & lens = 0.34$^{*}$ , host = 0.29 \\
        DDC6 & 71.3 $\pm$ 2.2 & 0.55 $\pm$ 0.01\\
        PDDEL1 & 41.6 $\pm$ 1.7 & 0.32 $\pm$ 0.02\\
        N10 & 24.9 $\pm 0.8$ & 0.14 $\pm$ 0.02\\
        GCD\_r51d4 & 20.1 $\pm 0.6$ & 0.19 $\pm$ 0.01\\
        SCH7p0 & 136.0 $\pm$ 13.0 & 0.85 $\pm$ 0.05\\
        M11\_05 & 22.5 $\pm$ 0.9 & 0.07 $\pm$ 0.02\\
        N150def & 46.1 $\pm$ 0.3 & 0.0 $\pm$ 0.01\\
        \hline
        $R_{V, \mathrm{host + lens}} = 3.1$\\
        DDC6 & 118.7 $\pm$ 5.0 & 0.54 $\pm$ 0.01\\
        PDDEL1 & 57.3 $\pm$ 3.2 & 0.32 $\pm$ 0.02\\
        N10 & 28.3 $\pm$ 1.2 & 0.13 $\pm$ 0.01\\
        GCD\_r51d4 & 24.3 $\pm 1.1$ & 0.18 $\pm$ 0.01\\
        SCH7p0 & 316.5 $\pm$ 13.6 & 0.87 $\pm$ 0.01\\
        M11\_05 & 25.0 $\pm$ 1.4 & 0.09 $\pm$ 0.02 \\
        N150def & 46.1 $\pm$ 0.3 & 0.0 $\pm$ 0.01\\
        \hline
	\end{tabular}
    }
    \begin{tablenotes}\footnotesize
\item $^{*}$ Magnification-weighted total $E(B-V)_{\mathrm{lens}}$.
\end{tablenotes}
\end{table}

\begin{figure}
    \centering
    \includegraphics[width=0.5\textwidth]{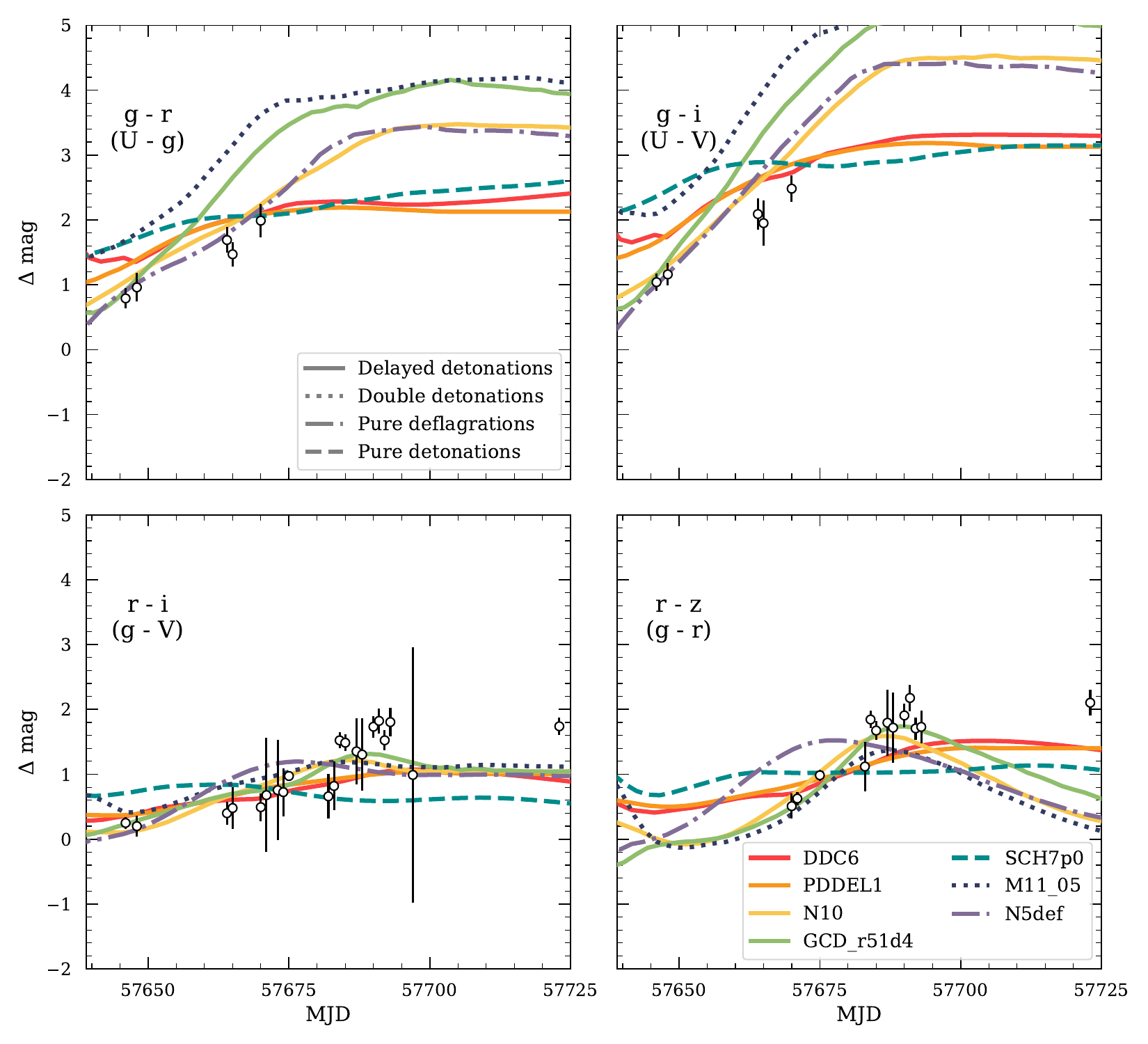}
    \caption{Colour curves of iPTF16geu compared to explosion models with best-fitting magnification and extinction parameters. Models are represented in the same way as in Figure \ref{fig:colourcurves}.}
    \label{fig:colours_2}
\end{figure}

The extinction and magnification values presented in Table~\ref{tab:magnification_dust} generally result in improved agreement with the observed light curves and colour evolution, as shown in Figs.~\ref{fig:lcs_2} and \ref{fig:colours_2}. With the fitted parameters, the N10 model shows significant improvement and provides the best overall match to the light curves and colours in all bands. The decrease in magnification results in the model generally matching the shape of the secondary maximum in the $z$-band. This model does however still evolve faster than iPTF16geu in the $g$ and $i$-bands. Moreover, Fig.~\ref{fig:spectra} shows that the spectral properties of this model are remarkably different to iPTF16geu. The PDDEL1 model, which showed greater similarities with the spectra, also generally matches the colour evolution, with an exception being the $r-z$ colour that shows a slower reddening over time. For the DDC6 model, the new magnification and extinction values do not solve the disagreement in the $g$-band. Overall our results show that while fitting for the magnification and extinction of each model can lead to improved agreement, it does not result in a model matching all features. Differences between models and observations highlighted by our comparisons therefore likely point to intrinsic differences between the two rather than extrinsic factors, such as the level of extinction assumed. 

\begin{figure}
    \centering
    \includegraphics[width=0.5\textwidth]{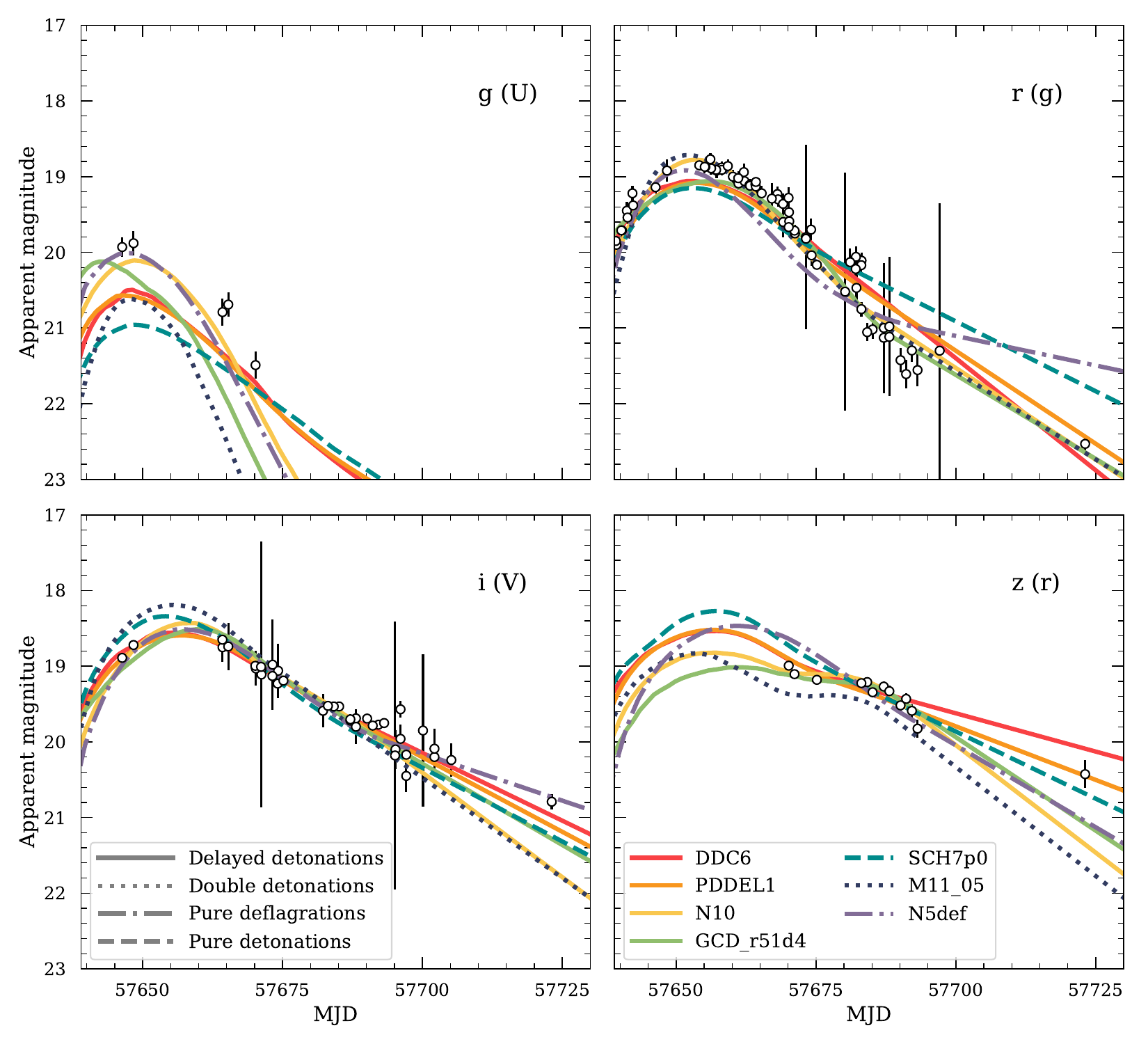}
    \caption{Light curves of iPTF16geu compared to explosion models with best-fitting magnification and extinction parameters (assuming $R_{V, \mathrm{lens + host}} = 2$). Models are represented in the same way as in Figure \ref{fig:norm_lc}.}
    \label{fig:lcs_2}
\end{figure}

The PDDEL1 and N10 models show generally improved agreement when fitting for magnification and extinction. Both models favour a lower amount of extinction compared to the \texttt{hsiao} template fit from \cite{Dhawan2019}, which could have a significant impact on $R_{\textrm{NUV}}$. We find that for the $R_{V} = 2$ case, applying the extinction correction from our PDDEL1 fits results in a $\sim$30\% decrease in $R_{\textrm{NUV}}$ that brings the early observations closer to the model. This does not however account for the discrepancy at later times. From our N10 fits, $R_{\textrm{NUV}}$ shows a similar decrease of nearly 50\%. For $R_{V} = 3.1$, the PDDEL1 and N10 extinction values result in a $\sim$40 -- 50\% decrease in $R_{\textrm{NUV}}$, again bringing the observations closer to the models, but not accounting for the potential late time flux excess. Therefore, as with the light curve and colour comparison, while the changes in extinction parameters results in improved agreement between the models and observations, these are not sufficient to fully explain differences in the UV flux ratio.

Independent measurements of extinction would be valuable in order to distinguish between models in the rest-frame optical bands and highlight intrinsic differences compared to the observations.

\section{Discussion}
\label{sec:discussion}

\subsection{Limitations}
In the following section, we discuss the limitations associated with our spectroscopic and photometric comparisons of iPTF16geu and explosion model predictions. 

Figures~\ref{fig:colourcurves} and \ref{fig:colours_2} show that none of our chosen models match the colour evolution post-maximum in all bands. Moreover, we find that the DDC6 and PDDEL1 models provide some similarities with observations of iPTF16geu, but are fainter in the NUV (observer-frame $g$-band), even after re-fitting for magnification and extinction parameters, and evolve more slowly in the $z$-band. These discrepancies could be related to the effects of microlensing, which is the additional (de)magnification from stars and other compact objects in the lens galaxy \citep{Dobler2006}. Microlensing has also been proposed as a possible explanation for the discrepancy in flux ratios between lens models and observations of iPTF16geu \citep{JMDiego22, Arendse25}. In our photometric analysis, we have applied the same magnification correction to all wavelengths. When the SN disk expands to a certain radius however, the microlensing effect becomes chromatic, affecting different wavelengths to varying degrees. This would impact the observed colour curves. The chromatic phase of microlensing typically begins on average 3 rest-frame weeks after the explosion, but in some cases it can start a few days after the explosion \citep{Huber19}. The median rest-frame time between explosion and peak $B$-band brightness for the models in this comparison is 17 days. Therefore, our observations are likely to be comparable to the start of the chromatic phase. \cite{Huber19} showed that in certain configurations microlensing would make colour curves appear bluer, which would be consistent with the observed colours of iPTF16geu. Chromatic microlensing could therefore be responsible for at least some of the discrepancies between iPTF16geu and the models presented here -- discrepancies may not solely be due to limitations of the models. Testing this however, would require constraining the position of the SN relative to the micro-caustics of the stars in the lens galaxy, which is beyond the scope of this paper. Alternatively, observations as early as possible, during the achromatic phase, would also reduce the impact of microlensing on model comparisons.

One assumption made in our analysis was the choice of reddening parameters ($A_V$ and $R_V$). For our analysis we chose the values of the fiducial model from \cite{Dhawan2019}, which assume the same $R_V$ for the host and the lens. In Section \ref{sec:new_dust_estimates} we further explored the impact of dust estimates by recalculating magnification and extinction parameters based on model fits to the data. We found that these updated fits improved agreement with the observed colours in some cases, such as with N10. We also tested our results for alternative choices of dust parameters presented by \cite{Dhawan2019}. We found that choosing a different set of extinction parameters (specifically $R_{V, \mathrm{lens}}$ = 1.8 and the higher $E(B-V)$ associated with it), results in the iPTF16geu light curves becoming approximately 0.3 mag bluer, but Fig.~\ref{fig:colourcurves} already shows that iPTF16geu is bluer than most models (especially in $g-r$). Moreover, the UV flux ratio increases as well, distancing it further from the models. On the other hand, setting both the $R_V$ of the host and the lens to be the same as the MW $R_V$ and using the corresponding $E(B-V)$ fit values from \cite{Dhawan2019} makes the SN even bluer and also shows a greater increase in the UV flux ratio, amplifying the discrepancies between the observations and the models. This highlights the importance of independent measurements of dust reddening to place more accurate constraints on explosion scenarios. 

\begin{figure}
    \centering
    \includegraphics[width=\linewidth]{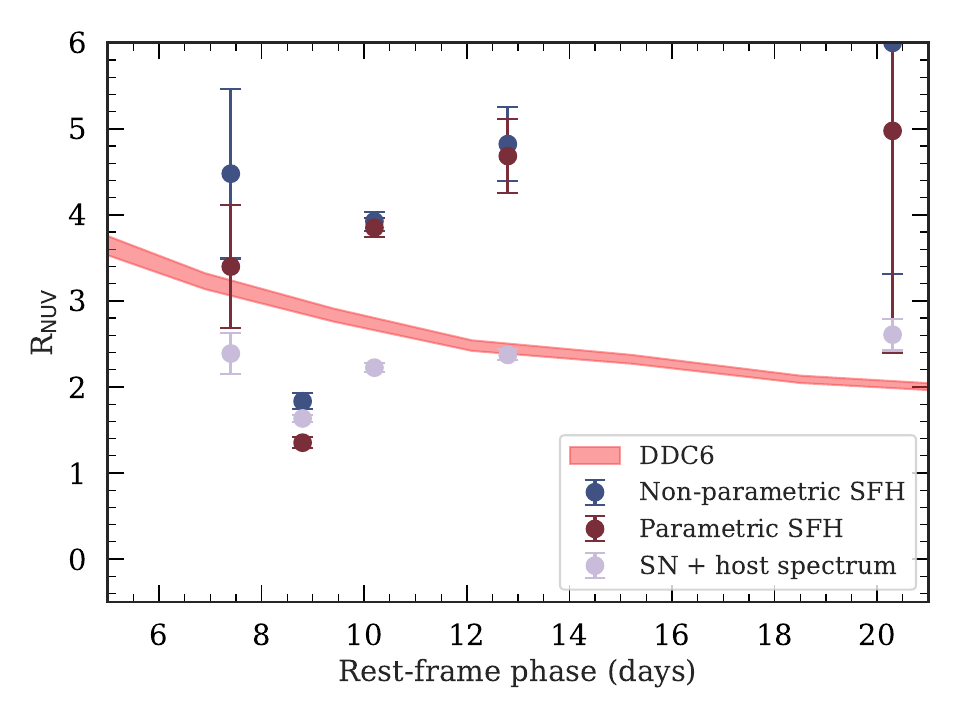}
        \caption{Dependence of UV ratio on the modelled galaxy spectrum. The choice of SFH used for modelling does not heavily impact UV ratios measured. Taking the measurements without host galaxy subtraction does result in a lower UV ratio.}
    \label{fig:SFH}
\end{figure}

The NUV region of the spectrum is particularly affected by the choice of host and lens galaxy subtraction. As detailed in Appendix \ref{appendix:galaxymodel}, we used a non-parametric star formation history (SFH) model for both the lens and the host galaxy in order to allow for flexibility in the fitting of the photometry. Alternatively, for the same lens galaxy \cite{Arendse25} use a parametric SFH. We therefore also fit our photometry with a delayed exponential function, which linearly increases at early times and exponentially decreases at later times, and find that the UV ratio remains mostly unchanged within the uncertainties (Fig.~\ref{fig:SFH}). Hence, our results are not strongly affected by the choice of SFH used in modelling the galaxy light. In Fig.~\ref{fig:SFH}, we also show the impact of not removing the galaxy contamination. This results in a flatter UV ratio up to phase +12.8 days post-maximum and a UV ratio evolution that does not match the best-fit model (DDC6), indicating that as the UV flux of the SN decreases, most of the flux comes from the lens galaxy light. Due to the large amount of contamination from the lens galaxy, \cite{Cano2018} showed the importance of subtracting the host and lens galaxy spectra in order to accurately measure the pEWs of SN features. We find that a similar conclusion applies to the UV ratios. 

In this work, we compared synthetic observables from models that were obtained through different simulation codes. Overall we found that the DDC6 and PDDEL1 models produced the best agreement, even compared to other delayed detonation models such as N10. This likely arises from multiple reasons, including differences in the explosion and radiative transfer setups of models (e.g. explosion codes, atomic datasets), the geometry and ejecta predicted by the models (DDC6 and PDDEL1 are 1D simulations of the explosions and radiative transfer whereas the N10 model is a fully three-dimensional simulation), and the radiative transfer codes used (DDC6 and PDDEL1 use the full NLTE approach of \textsc{cmfgen} whereas the N10 model uses the \textsc{artis} NLTE approximation). \cite{Blondin2022} demonstrate how using different radiative transfer codes for the same model impacts the resulting light curves and spectra. They present comparisons for the DDC10 and DDC25 models, as well as two toy models, and find that for the same model \textsc{artis} produces overall slower pseudo-bolometric rise times and slightly fainter pseudo-bolometric light curves. They present the individual $UBVRIJHK$ light curves for one of their toy models and show that the biggest differences are in the $IJHK$-bands, which we do not include in our comparison. For the bands in this work the biggest difference is in the $U$-band, which is 0.7 mag fainter in the \textsc{artis} simulations. Resimulating \textsc{artis} models with \textsc{cmfgen} instead would still not be sufficient to match the observed $g$-band light curve (which corresponds to the rest-frame $U$-band). The $R$-band in the \textsc{artis} results shows a slower rise time and faster decay, and is brighter than the \textsc{cmfgen} results by $\approx 0.4$ mag, which falls within the uncertainties of the $z$-band light curve. The $BV$-bands show differences of $\lesssim0.2$ mag between \textsc{cmfgen} and \textsc{artis} simulations, with the \textsc{artis} simulations showing a faster decay rate but slower rise time. A slower rise time would better align the DDC6 and PDDEL1 light curves with observations. The differences in spectral shapes could also be partially explained by the choice of modelling code, as spectra generated with \textsc{cmfgen} result in higher mean ionization, independently of the ejecta velocity, leading to different spectral shapes. We conclude however that changes in the explosion calculation and the radiative transfer treatment would not strongly affect our conclusions regarding the mismatch between the $g$-band and observations, but make it challenging to fully discriminate between explosion scenarios.

Here we considered results based on either fully one-dimensional simulations (hydrodynamics and radiative transfer) or angle-averaged spectra derived from three-dimensional simulations, therefore ignoring any viewing angle dependencies. \cite{Maeda2010} suggested that viewing angle dependencies could be the origin of the diversity of SNe~Ia spectral evolution. The viewing angle dependence has been studied in detail for double detonation models. \cite{doubledetonations2021} and \cite{collins22} found that the light curves showed a strong angle dependence, especially in the $U$- and $B$-bands, where the viewing angle can vary peak brightness by up to 3 mags. This is due to a combination of line blanketing effects and the distribution of the model ejecta. The angle dependence decreases at redder bands since the optical depth decreases with wavelength. \cite{Collins2025} showed that the differences with viewing angle in double detonation models are strongly reduced when using a full NLTE treatment, thus suggesting that viewing angle dependencies are more likely insufficient to justify differences with the models. \cite{Sim2013} also show some angle dependence in their delayed detonation models, which result in up to 0.3 mag. uncertainty in the light curves in the bluer bands, which reduces to $\approx$ 0.1 mag in the other bands. The pure deflagrations also exhibit some angle dependence, though less significant \citep{fink--14}.

\subsection{Comparison with models}
As discussed by \cite{Johansson2020}, iPTF16geu exhibits spectral properties comparable to other normal SNe~Ia. Based on the magnification and dust corrections presented by \cite{Dhawan2019} however, iPTF16geu is slightly over-luminous at blue wavelengths. A viable explosion model should be able to reproduce these properties.

In Section \ref{sec:photometry}, we present light curve and colour comparisons of iPTF16geu and our models, finding that the DDC6 and PDDEL1 models exhibit the strongest matches to the light curves and colours of iPTF16geu. A slight offset remains at early times in the $gr$-bands, but, as suggested by \cite{Dessart14}, adjustments to the setup for the pulsational phase could bring the observations and the PDDEL1 model into closer agreement. Introducing a $\approx$1 day time shift in the models (see e.g. \citealt{Mazzali14}) improves agreement with the $g$-band rise time without impacting the $iz$ light curves, but the mismatch in the $g$-band magnitudes and $r$-band rise times remains after applying this shift. Moreover, this shift would make the discrepancies in the $g-r$ colour more prominent, as the SN would appear redder.

Spectroscopically, these models provide a good overall match to the SN (Fig.~\ref{fig:spectra}). Up to $\approx23.8$ days, the observed spectra appear to match the optical colours of the models. The widths of different features, such as the \ion{Si}{ii} and the \ion{Ca}{ii} H\&K, are also reproduced by the models. The PDDEL1 model displays similar \ion{Si}{ii} velocities at the earliest phases for which we have data, but a slower velocity gradient evolution for this feature, whereas the \ion{Si}{ii} is overpredicted by the DDC6 model. \cite{blondin--13} find their models to be systematically blueshifted compared to the normal SN~Ia subclass to which iPTF16geu belongs according to \cite{Johansson2020}. They attribute this effect to an excess of kinetic energy in the ejecta for these models. This effect is also observed in other features such as the \ion{Ca}{ii} feature, which displays a higher velocity in the DDC6 model despite having similar widths. Both of these models show relatively similar $R_{\mathrm{NUV}}$ values at the phases covered by the observations, as seen in Fig.~\ref{fig:UVflux}. Figure~\ref{fig:UVflux} also shows that the PDDEL1 model is bluer than the DDC6 model at early times, which is a result of the hydrodynamic interaction that arises in this scenario. We note however that while both of these models are calculated using the NLTE treatment, they are 1D explosions. As discussed in Section \ref{sec:models}, multi-dimensional effects impact the ejecta structure. It remains unclear whether three-dimensional simulations of the DDC6 and PDDEL1 models would still provide the best match to iPTF16geu.

We also include the N10 delayed detonation model presented by \cite{Seitenzahl2013} and \cite{Sim2013}. This model matches the $riz$ light curves but fails to reproduce the $g$-band light curve and the blue colours of iPTF16geu when allowing the level of extinction to vary. We consider the wider range of DDT models presented by \cite{Sim2013}, but find that all appear too red for iPTF16geu. 
Spectroscopically, the N10 model does not reproduce the overall spectral shape or the \ion{Si}{ii} feature observed in iPTF16geu. \citet{Sim2013} demonstrate that models with a larger number of deflagration sparks exhibit a stronger \ion{Si}{ii} feature and could therefore provide a better spectral match. Such models are however intrinsically fainter and generally redder. As shown in Section~\ref{sec:new_dust_estimates}, these discrepancies could be resolved if the magnification were higher and different extinction corrections were applied. Independent constraints on magnification and extinction would be valuable to fully assess whether DDT models offer a plausible explanation for iPTF16geu.

The pure detonation model (SCH7p0; \citealt{Blondin17}) also matches some of the properties of iPTF16geu, more specifically at early (<20 days) phases, and produces spectra comparable to our delayed detonation models. This results from burning in the outer ejecta layers occurring at similar densities for both these sub$M_{\mathrm{Ch}}$ and $M_{\mathrm{Ch}}$ models, producing similar outer ejecta profiles. The lower WD density however, leads to a more limited production of neutron-rich, stable IGE isotopes in the inner ejecta. This is evident in the much bluer colours of the SN at later phases observed in Fig.~\ref{fig:colourcurves}. This effect is also observable in Fig.~\ref{fig:UVzoom}, as the UV flux is brighter and shows more distinguishable features such as \ion{Fe}{ii}, which are not observed in iPTF16geu. Around $\approx$2 months post-maximum, the spectra of iPTF16geu show significant differences compared to the SCH7p0 model, suggesting that the structure of the inner ejecta also deviates significantly. The structure predicted by the delayed detonation models may provide a closer match to that produced by iPTF16geu. In this work we did not explore alternative versions of the double detonation scenario relying on less massive He shells (such as the Dynamically-Driven Double-Degenerate Double-Detonation; \citealt{Tanikawa18}). The observables in these models are less impacted by the burning products of the He shell \citep{Pollin24} and therefore cannot be ruled out as a possible scenario.

Despite matching some light curve properties, the M11\_05 model and the N150def model show quite different spectral properties. We find that both models exhibit significant line blanketing towards bluer wavelengths, which also results in redder colours than those observed for iPTF16geu. In the case of double detonations, this is due to line blanketing caused by IGE in the outer layers of the ejecta that are synthesised in the helium detonation \citep{collins22}. Some line blanketing is also expected for the pure deflagrations, as the turbulent burning leads to an almost completely mixed ejecta structure, and therefore a larger fraction of IGE in the ejecta compared with other explosion models. The spectra of iPTF16geu do not show easily identifiable features beyond $\approx \ 3\,500$ \AA, which could be a hint of some amount of line blanketing. The blue colours however, indicate that the lack of identifiable features may instead be due to a low signal-to-noise ratio or contamination from the galaxy subtraction. Both the strong line blanketing and the presence of \ion{Ti}{ii} are a general prediction of the double detonations with thick helium shells presented by \cite{doubledetonations2021}, therefore thick helium shell detonations can likely be ruled out as producing iPTF16geu.

Based on the available data, we find that the DDC6 and PDDEL1 produce the best qualitative agreement with observations of iPTF16geu. Overall, both models yield a large Ni$^{56}$ mass and predict similar observable features after maximum light. Before reaching maximum bolometric light they show significantly different spectroscopic properties, arising from differences in their ejecta structures. For example, the PDDEL models predict the presence of \ion{C}{ii} absorption during the first approximately one week after explosion. Earlier observations would aid in discriminating between these scenarios for future glSNe.

\subsection{Comparison with other SNe~Ia}

As mentioned, none of the models are able to fully reproduce the properties of iPTF16geu. We include comparisons with other SNe~Ia to determine if these discrepancies are due to limitations in the physics of the models or if they indicate hints of an evolution in SN properties with redshift that are not accounted for. The optical spectra of SN~2011fe show similar spectral shapes and feature strengths as iPTF16geu, although with a lower \ion{Si}{ii} and \ion{Ca}{ii} H\&K velocities, suggesting this explosion had lower kinetic energies. We also find that the UV flux ratio at phases $t \in [8,13]$ is consistent with the observed ratios for SN~2011fe. \cite{Dessart14} find one of their PDDEL models (PDDEL4) to be a good physical model for SN~2011fe, while \cite{Ropke12} find some preference for a violent merger scenario. Both of these studies find disagreement between models and observations at early times. We find a similarly good agreement between the spectra of SN~2011fe and the DDC6 and PDDEL1 models from peak to $\approx$3 weeks post-maximum. At later phases, although the models still roughly reproduce the colours of this SN, the spectrum of SN~2011fe is more similar to that of iPTF16geu, and we find some discrepancies between the models and observations.
    
We also include in our comparison the UV spectra of two normal SNe~Ia at redshifts $z\sim 0.4$ and find some level of agreement in the spectra. Comparing with the models, we find that SNLS-03D1ar at +5.3 days appears to match the PDDEL1 and N10 models best, although it exhibits slightly lower ejecta velocities. On the other hand, SNLS-05D4ay (+8.2 days) shows a higher \ion{Ca}{ii} H\&K velocity. Comparing to PS1-10afx, we find that PS1-10afx displays a \ion{Fe}{ii} 3\,250~\AA\ feature that is present in the pure detonation models and, with less intensity, in some of the delayed detonation models, but does not appear in the spectrum of iPTF16geu. We also measure a greater UV flux for PS1-10afx, but this is most likely explained by its earlier phase and the rapid evolution of the UV spectrum. In general, the DDC6, PDDEL1, and N10 models provide a good match to the shape of the UV spectrum of this SN. As shown in Fig. \ref{fig:UVflux}, the UV ratio measurements of PS1-10afx show good agreement with the DDC6 model estimates. Overall, the small variations in spectral features across this SN sample can be explained by our range of explosion models. The lack of variation in model preference across the SN redshift range supports the conclusion that SN properties show no significant evolution with redshift, which is consistent with previous studies \citep{Hook2005, Petrushevska17, Dhawan2024}. Larger samples and higher quality observations however are required to determine the statistical significance of these trends.

\section{Conclusions}
\label{sec:conclusions} 
In this paper, we analysed observations of iPTF16geu, a gravitationally lensed SN~Ia at $z = 0.409$, and compared them against predictions from theoretical explosion models. Models were selected from the literature and cover a wide variety of proposed scenarios, including pure detonations, delayed detonations, double detonations, and pure deflagrations. 

Overall, we found a preference for the delayed detonation scenarios; particularly the DDC6, PDDEL1 and N10 models provide the best qualitative agreement with observations of iPTF16geu during the early phases. The DDC6 and PDDEL1 models show the most similarities with the observed light curves, closely matching the $r$- and $i$-band light curves, the $z$-band rise times, and the $g$-band decline rates. The N10 model also shows good agreement with the data, but requires lower magnification and extinction parameters than previously reported assuming a SN~Ia template fit. In addition, the DDC6 and PDDEL1 models generally match the overall shapes, strengths, and velocities of many spectral features. None of the models however can match all observed features. The models also generally reproduce the shape of the evolution of the flux in the NUV regions but show an overall lower NUV flux ratio. We explored the impact of host galaxy subtraction on the UV evolution and found that while removal of host and lens galaxy spectra is necessary, the choice of host template fitting does not significantly affect our results. 

Comparing our observations of iPTF16geu to other SNe~Ia, we found no evidence in favour of a systematic redshift evolution in their spectral properties, consistent with other studies \citep{Petrushevska17, Dhawan2024}. The UV flux of iPTF16geu shows indications of an excess compared to models and other low-$z$ SNe~Ia, but the low signal-to-noise ratios of the observations at late times mean we are unable to determine whether this is significant. We note that \cite{Maguire2012} found a similar excess in the UV flux of a sample of intermediate redshift ($0.4 \leq z \leq 0.9$) SNe, but this sample did not appear to introduce a significant cosmological bias in the Hubble diagram. The significance of this excess is also strongly dependent on the choice of extinction parameters.

Based on our comparisons, the DDC6 and PDDEL1 delayed detonation models provide the best overall agreement with iPTF16geu, but the multiple limitations concerning the explosion set up and radiative transfer treatment make it challenging to reach unbiased conclusions on the explosion scenario. Nevertheless, our analysis highlights that obtaining observations as early as possible is crucial for discriminating between models, particularly for glSNe that may be affected to some extent by chromatic microlensing. Strong lensing time-delays can however simplify this task. In systems with sufficiently long time delays, one can predict the appearance of the additional images and trigger follow-up that would allow for observations at these earlier phases to be obtained. LSST will find hundreds of glSNe over its 10-year survey \citep{Goldstein19, Arendse24, SainzdeMurieta23, SainzdeMurieta24}. \cite{FoxleyMarrable20} showed that a large fraction of these could be detected before the appearance of the system's last image. This work highlights the importance of accurate lens modelling in predicting the appearance of multiple images to trigger spectroscopic follow-up at the most constraining phases. This would leverage the properties of glSNe to obtain further insight into the intermediate redshift SN~Ia population.

\section*{Acknowledgements}
The authors would like to thank Stuart A. Sim, Christine E. Collins and Fionnt\'an P. Callan for sharing some of the explosion models, as well as providing valuable comments on the draft. They would also like to thank Sophie Newman for useful discussions on galaxy SED fitting and Luke Weisenbach for helpful comments on microlensing. This project has received funding from the European Research Council (ERC)
under the European Union’s Horizon 2020 research and innovation
programme (LensEra: grant agreement No 945536).

T.~E.~C. is funded by a Royal Society University Research Fellowship. MRM acknowledges a Warwick Astrophysics prize post-doctoral fellowship made possible thanks to a generous philanthropic donation.

\section*{Data Availability}
This work made use of the Heidelberg Supernova Model Archive (HESMA), \url{https://hesma.h-its.org}. The iPTF16geu spectra are available on WiseRep.



\bibliographystyle{mnras}
\bibliography{lensedSN} 


\appendix
\section{SED modelling for galaxy subtraction}
\label{appendix:galaxymodel}

The P200 spectrum of iPTF16geu after fading does not extend far enough into the UV region.
Thus, we construct model SEDs of both the lens and host galaxy from matching star-formation history (SFH) models to broad band photometric data. \cite{Cano2018} show the importance of subtracting the contributions from the galaxies from the observed spectra for measuring the SN intrinsic properties. As they expect the contribution of the host-galaxy to be negligible, they match the galaxy spectra to an elliptical galaxy template by \cite{Manucci2001} and use the "colour-matching" technique outlined by \cite{Foley2012} to ensure the colours of the spectrum match the photometry. \cite{Johansson2020} use a similar method and the same set of galaxy templates, but combine the host and lens galaxy. In this work we allow for more freedom in the shape of the galaxy SEDs by fitting the parameters of the SFH to the photometry. This only requires a prior on the SFH function, removing assumptions about the shape of the SED.

\begin{figure*}
    \centering
    \includegraphics[width=\textwidth]{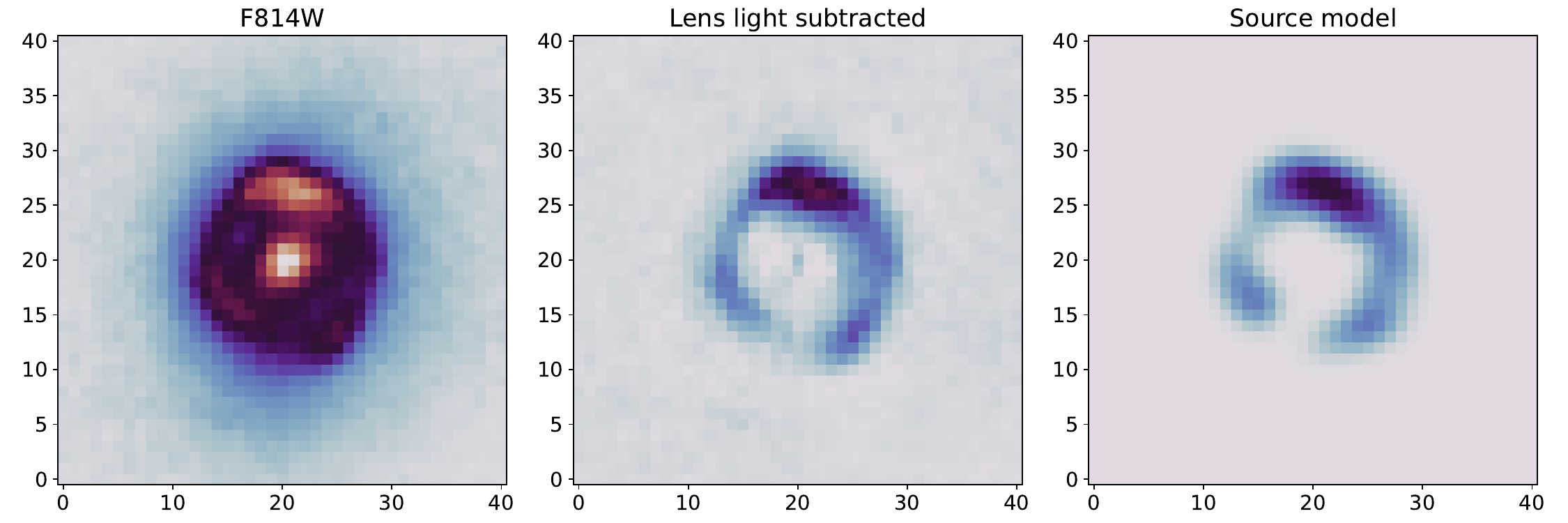}
    \caption{Example of lens light subtraction on the F814W filter image.}
    \label{fig:lightsubtraction}
\end{figure*}

We use public data of the lensed system from GALEX ($NUV$ filter), SDSS ($ugriz$ filters), Pan-Starrs ($grizy$ filters), and 2MASS ($JHK$ filters). We calculate the fraction of light corresponding to the lens and host galaxies in HST template images ($F390W$, $F475W$, $F625W$, $F814W$, $F110W$, $F160W$ filters) taken on the 10th November 2018, after the SN had faded. For the optical ($F390W$, $F475W$, $F625W$, $F814W$) filters, we reconstruct the observed image by constructing a simple lens model and separate the lens light and the light from the arcs, as shown in Figure~\ref{fig:lightsubtraction}. The lensed arcs are not distinguishable in the IR ($F110W$ and $F160W$) data, making it hard to constrain our lens model. In these filters, we attempt to subtract the light of the lens by fitting a Sersic profile to the lens light and subtracting it. This provides a good approximation for the photometry, although with larger uncertainties, which we then reflect in our measurements. We then perform aperture photometry on the light subtracted images to calculate the fractional flux coming from the lens and the source in an aperture of radius 0.4''.
We use Gaussian Processes (GP) to extrapolate the fractional fluxes beyond the HST wavelengths, by determining the fraction of light coming from the lens (and source) as a function of wavelength. This fraction is then used to determine the photometry of the lens galaxy and the source galaxy.

\begin{figure*}
    \centering
    \includegraphics[width=\textwidth]{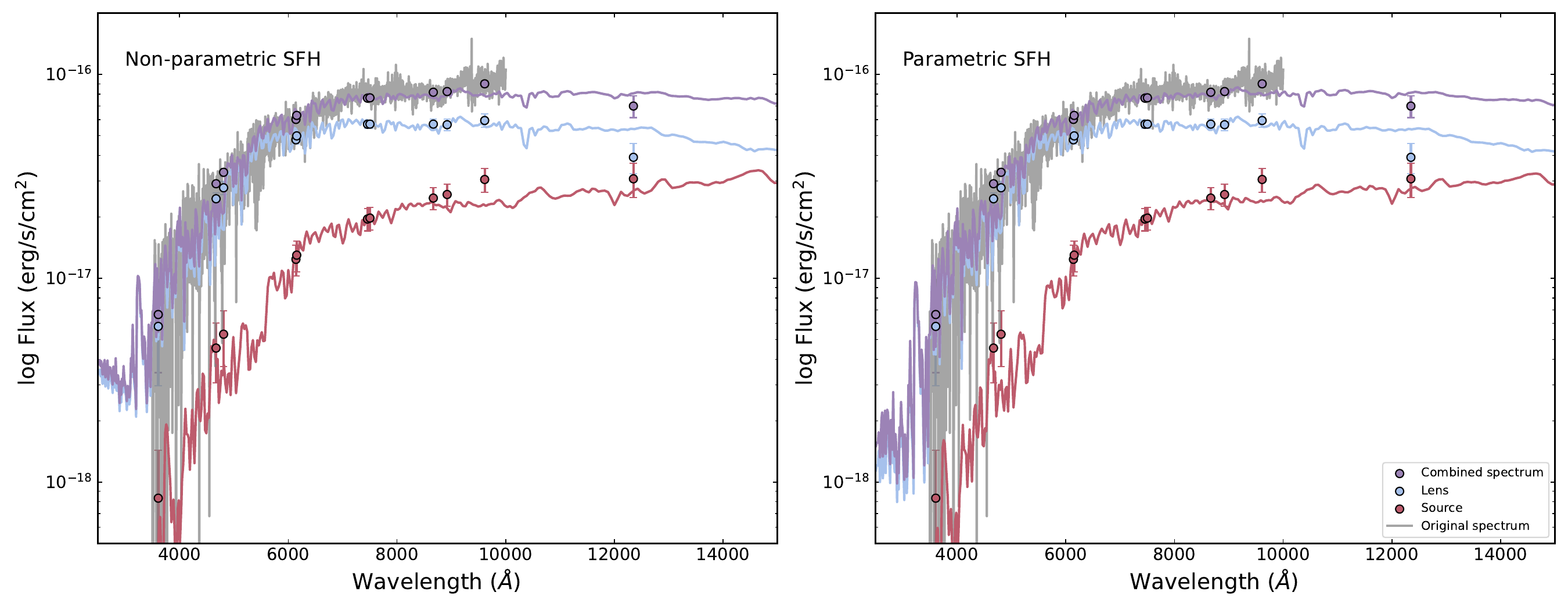}
    \caption{Best-model fits from \texttt{BAGPIPES} for both our non-parametric and parametric models.}
    \label{fig:bagpipes_fit}
\end{figure*}

We model the spectrum of the lens and the source galaxy using \textsc{bagpipes} \citep{BAGPIPES}, with the photometry we calculated. As we mentioned earlier, this only requires a prior on the shape of the SFH. Due to their increased flexibility in comparison to parametric SFH models, we use the non-parametric models by \cite{Leja2019} with a continuity prior on the star-formation rate (SFR) over time, which fits directly for $\Delta$log(SFR) between adjacent time bins. This prior explicitly weights against sharp transitions in the star formation rate and is easier to tune in order to follow expectations from theoretical models of galaxy formation. We also include a dust parameter following the Cardelli extinction law \citep{Cardelli89}. We fix the redshift of the galaxies, the $A_V$ based on \cite{Dhawan2019}, and provide uniform priors for the other parameters. We use a tight prior on the mass of the lens galaxy by using the relation between mass and velocity dispersion from \cite{Bernardi10}. The resulting best-fit models for both SFH are shown in Figure~\ref{fig:bagpipes_fit}. We compare our fits with both the photometry and original observed host galaxy spectrum, showing our model provides a good fit to our estimated photometry in the UV-optical bands, and provides a close fit to the data in the IR, where the data is less constraining. We also find that our model associated to the parametric SFH has a steeper flux increase in the UV, but this does not affect the UV flux ratios considerably, as seen in Figure~\ref{fig:SFH}.

\section{Summary of explosion models}
\label{appendix:models}
\begin{table*}
	\centering
	\caption{Summary of properties of explosion models used in the spectral comparison. These properties correspond to the light curves at $z = 0.409$}
	\label{tab:models}
    \resizebox{\textwidth}{!}{
	\begin{tabular}{llccccccccr}
		\hline
    		 Name & & Source &$M_g$ & $M_r$ & $M_i$ & $M_z$ & $\Delta m_{15}^g$ & $\Delta m_{15}^r$ & $\Delta m_{15}^i$ & $\Delta m_{15}^z$\\
         \hline
         \textbf{iPTF16geu} & & & \textbf{$-20.0\pm0.2$} & \textbf{$-20.2\pm0.2$} & \textbf{$-19.9\pm0.2$}& \textbf{$-19.8\pm0.4$}& \textbf{$0.71\pm0.16$}& \textbf{$0.60\pm0.11$} & \textbf{$0.44\pm0.10$}& \textbf{$0.60\pm0.07$}\\
         \hline
         \textbf{Delayed detonations}&DDC6& B13&  $-$19.3 & $-$20.0 & $-$19.9 & $-$19.7 &0.78 & 0.49 &0.48& 0.41\\
         &PDDEL1& D14&$-$19.2 &$-$20.0 & $-$20.0 & $-$19.8 &0.73 &0.53 & 0.37 & 0.39\\
         &N10& S13&$-$19.0& $-$20.2& $-$20.3& $-$19.9& 0.98& 0.84& 0.69&0.25\\
         &GCD\_r51d4 & L22 & $-$19.4 & $-$20.2 &$-$20.6 & $-$19.9 & 0.87 & 0.65 & 0.61 & 0.19	\\
         \hline

        \textbf{Detonations}&SCH7p0& B17 &$-$19.4& $-$20.2& $-$20.1&$-$19.8 &0.59 &0.49 &0.61 &0.54\\
         \hline
        \textbf{Double detonations}&M11\_05& G21 &$-$18.3 & $-$20.1 & $-$20.5 & $-$19.9 &1.58 & 0.83 & 0.63 & 0.51
\\
         \hline
         \textbf{Deflagrations}&N150def& F14& $-$17.9 & $-$18.9 & $-$19.29& $-$19.3&1.07 &0.81 & 0.49 &0.42\\
         \hline
	\end{tabular}
    }
    \begin{tablenotes}\footnotesize
\item B17: \cite{Blondin17}, G21: \cite{doubledetonations2021}, S13: \cite{Sim2013}, B13: \cite{blondin--13}, D14: \cite{Dessart14}, L22: \cite{Lach22GCD}, F14: \cite{fink--14}
\end{tablenotes}
\end{table*}

\section{Light curves for $R_{V} = 3.1$}
\label{appendix:rv}
\begin{figure}
    \centering
    \includegraphics[width=0.5\textwidth]{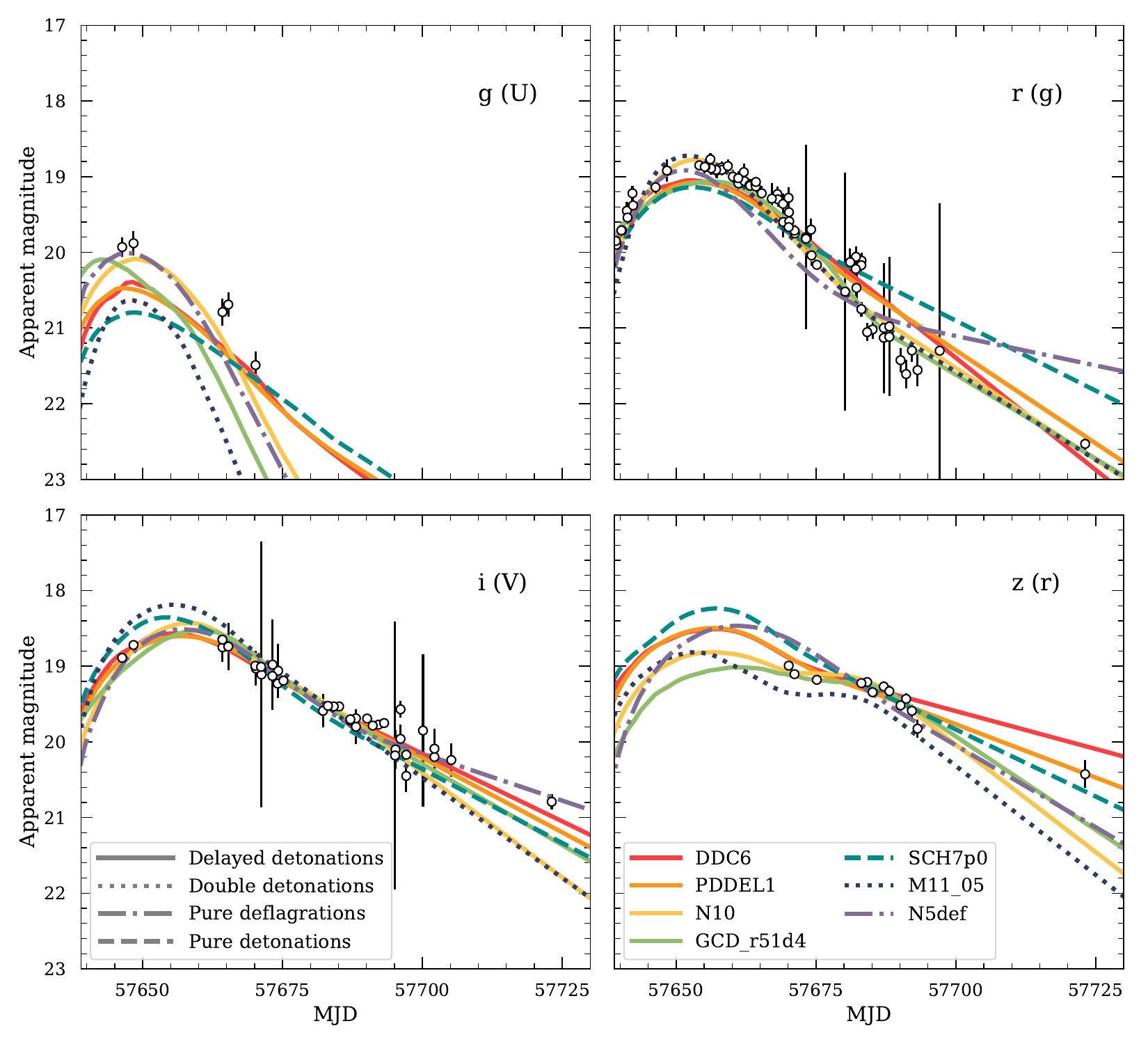}
    \caption{Same as Figure \ref{fig:lcs_2} but with $R_{V,\ \mathrm{host+lens}} = 3.1$}
    \label{fig:lcs_3}
\end{figure}

\begin{figure}
    \centering
    \includegraphics[width=0.5\textwidth]{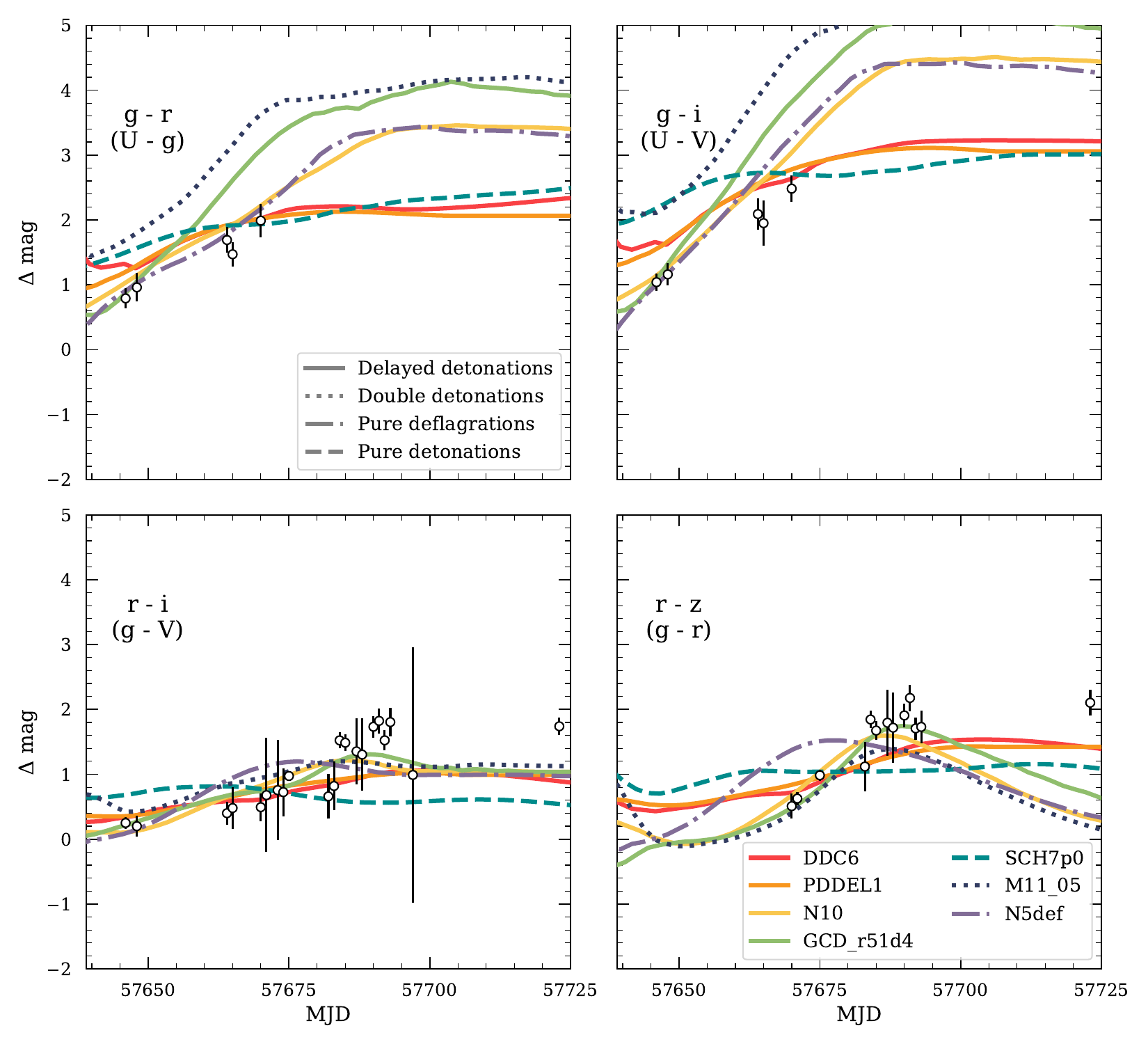}
    \caption{Same as Figure \ref{fig:colours_2} but with $R_{V,\ \mathrm{host+lens}} = 3.1$}
    \label{fig:colours_3}
\end{figure}

\bsp	
\label{lastpage}
\end{document}